\documentclass[journal]{IEEEtran}
\usepackage{amsmath}
\usepackage{amsthm}
\usepackage{amsfonts}
\usepackage{graphicx}
\usepackage{epstopdf}
\usepackage{color}
\usepackage[caption=false,font=footnotesize]{subfig}
\usepackage{subeqnarray}
\usepackage{algorithmic}
\usepackage{algorithm}
\usepackage{indentfirst,comment} 
\usepackage{cite}
\usepackage{tikz}
\usepackage{arydshln}
\usepackage{multirow}
\usepackage{booktabs}
\usepackage{slashbox}
\usepackage{mathtools}

\DeclareMathOperator*{\argmin}{arg\,min}

\newcommand{\bignorm}[1]{\left\lVert#1\right\rVert}
\newcommand{\norm}[1]{\lVert#1\rVert}
\newcommand{\blkdiag}[1]{\text{blkdiag}(#1)}
\newtheorem{mydef}{Definition}

\newtheorem{thm}{Theorem}
\newtheorem{rem}{Remark}
\newtheorem{lma}{Lemma}

\begin{document}
\title{An Adaptive Partial Sensitivity Updating Scheme for Fast Nonlinear Model Predictive Control
}

\author{Yutao~Chen, Mattia~Bruschetta, Davide~Cuccato, and Alessandro~Beghi%
\thanks{Yutao~Chen, Mattia~Bruschetta, Davide~Cuccato, and Alessandro~Beghi are with the Department of Information Engineering, University of Padova, Via Gradenigo 6/B, Padova 35131, Italy. Email: yutao.chen@dei.unipd.it; mattia.bruschetta@dei.unipd.it; davide.cuccato@dei.unipd.it; beghi@dei.unipd.it. Tel: +390498277626. Fax: +390498277699}}%

\maketitle
\begin{abstract}
In recent years, efficient optimization algorithms for Nonlinear Model Predictive Control (NMPC) have been proposed, that significantly reduce the on-line computational time. In particular, direct multiple shooting and Sequential Quadratic Programming (SQP) are used to efficiently solve Nonlinear Programming (NLP) problems arising from continuous-time NMPC applications. One of the computationally demanding steps for on-line optimization is the computation of sensitivities of the nonlinear dynamics at every sampling instant, especially for systems of large dimensions, strong stiffness, and when using long prediction horizons. In this paper, within the algorithmic framework of the Real-Time Iteration (RTI) scheme based on multiple shooting, an inexact sensitivity updating scheme is proposed, that performs a partial update of the Jacobian of the constraints in the NLP. Such update is triggered by using a Curvature-like Measure of Nonlinearity (CMoN), so that only sensitivities exhibiting highly nonlinear behaviour are updated, thus adapting to system operating conditions and possibly reducing the computational burden. An advanced tuning strategy for the updating scheme is provided to automatically determine the number of sensitivities being updated, with a guaranteed bounded error on the Quadratic Programming (QP) solution. Numerical and control performance of the scheme is evaluated by means of two simulation examples performed on a dedicated implementation.
Local convergence analysis is also presented  and a tunable convergence rate is proven, when applied to the SQP method.

\end{abstract}


\section{Introduction}
Nonlinear Model Predictive Control (NMPC) has been studied and applied intensively in the last decades. In NMPC, a nonlinear Optimal Control Problem (OCP) has to be solved on-line at every sampling instant. The OCP can be converted to a finite dimensional Nonlinear Programming (NLP) problem by direct methods, such as direct multiple shooting \cite{bock1984multiple} and direct collocation \cite{biegler2010nonlinear}. The NLP problem can then be solved by a number of optimization algorithms, e.g., Interior Point Methods (IPM) \cite{biegler2010nonlinear} and Sequential Quadratic Programming (SQP) \cite{powell1978fast}. Fast NMPC algorithms based on direct methods have been proposed to speed up  on-line optimization, see \cite{diehl2001real,zavala2009advanced, graichen2012real}.

Efficient SQP algorithms based on direct multiple shooting for systems governed by Differential Algebraic Equations (DAE) have been thoroughly studied  (see e.g. \cite{leineweber1999efficient}). One of the computationally demanding steps of SQP methods when applied to NMPC is the computation of sensitivities at each sampling instant, i.e. the Hessian of the Lagrangian and the Jacobian of the constraints. There are several methods for computing such sensitivities, e.g. finite difference \cite{nocedal2006numerical}, complex-step differentiation \cite{martins2003complex}, and automatic differentiation \cite{rall1981automatic}. 

Particularly, the Jacobian of constraints contains sensitivities of integration operators that parameterize continuous-time dynamics. Although efficient implementations of numerical integration with sensitivity generation are available \cite{kuhl2007muscod, houska2011auto}, sensitivity computation of this type still largely contributes to the overall on-line computational burden, especially for systems that are highly stiff or governed by implicit differential equations and DAEs.


In this paper, on one most promising SQP-based NMPC algorithms that is the Real-Time Iteration (RTI) \cite{diehl2002real}, in which only one SQP iteration is performed at each sampling instant, is taken as the reference approach. 
The underlying idea is to initialize the new NLP by using information from the previous one, including states, controls, and multipliers making the closed-loop trajectory converging as system dynamics evolve, i.e. ``on the fly'' \cite{gros2016linear}. 

In the RTI framework with multiple shooting parameterization, a number of tailored approaches are available that employ suitable inexact sensitivities. In Multi-Level RTI (ML-RTI) \cite{bock2007constrained}, sensitivities are updated every $m>1$ sampling instants. Hence, sensitivities are updated at a slower rate than other QP components. However, the choice of $m$ is not intuitive and generally application dependent, thus requiring a long and complex tuning procedure.
In ADJoint sensitivity RTI (ADJ-RTI)  sensitivities computed off-line are used \cite{wirsching2006fast, wirsching2008adjoint, kirches2012efficient, zanelli2016efficient}, and an adjoint sensitivity on-line computation is performed to identify the correct active-set and to ensure local convergence. Although computational cost is considerably reduced, thanks to the reduced number of sensitivity computations and condensing operations \cite{kirches2012efficient}, this approach is effective only for systems exhibiting mild nonlinearities. Recently, partial sensitivity updating schemes called CMoN-RTI and DOPUS, that are tailored for multiple-shooting based NMPC, have been proposed \cite{chen2017fast, chen2017inexact, van2017towards}. In such schemes, the multi-stage feature of NLP problems arising in NMPC applications and the iterative nature of the solver are exploited. As a result, sensitivities are partially updated between two consecutive sampling instants. A so-called Curvature-like Measure of Nonlinearity (CMoN) or norm-criterion is used in a monitoring strategy to decide which and how many sensitivities should be updated. However, these monitoring strategies rely on heuristics and are strongly dependent on the application at hand.

In this paper, the partial sensitivity scheme CMoN-RTI of \cite{chen2017inexact} is extended and improved. In particular, three main features are provided, namely: 
\begin{itemize}
    \item a solution accuracy control strategy;
    \item a practical tuning procedure;
    \item convergence analysis.
\end{itemize}
From parametric optimization theory, the accuracy of the QP solution is related to parameters in the monitoring strategy. An advanced tuning strategy for CMoN-RTI is here developed that provides an automatic way to select which and how many sensitivities should be updated, while guaranteeing the QP solution a bounded \emph{Distance to Optimum} (DtO) (i.e. the distance between the solution of the inexact sensitivity QP and that of the exact sensitivity QP). The tuning parameter is therefore the DtO tolerance, which has an important physical meaning.

The proposed scheme can significantly reduce the computational load when the system nonlinear dynamics are excited only on a small part of the prediction horizon (e.g., when regulating a system around its steady state or tracking a reference with look ahead).
Moreover, since the additional computational time required by CMoN-RTI with respect to RTI is almost negligible, CMoN-RTI is a sensible alternative in all the scenarios where RTI is effective, as it usually yields an improvement in the average computational performance, hence saving computational power, and possibly an increase of the control frequency. In the worst case, CMoN-RTI degrades to RTI. A practical implementation of the scheme is given and its effectiveness is demonstrated by closed-loop simulations on two classical examples. An application of CMoN-RTI applied to the SQP framework with multiple iterations is also introduced. A tunable local convergence rate is proven.

The paper is organized as follows. In Section \ref{sec2}, RTI and some inexact sensitivity schemes are briefly introduced to define the algorithmic framework. In Section \ref{sec3}, the CMoN-RTI scheme is presented in detail. Section \ref{sec4} is devoted to the derivation of the advanced tuning strategy and to practical implementation aspects. In Section \ref{sec5}, closed-loop simulation results using CMoN-RTI are shown. The CMoN-SQP is described in Section \ref{sec6}, and its convergence properties are discussed and demonstrated by a numerical example.

\section{Algorithmic Framework}\label{sec2}
In this section, the standard RTI scheme \cite{diehl2002real} is introduced as the algorithmic framework of the paper. ML-RTI  \cite{bock2007constrained} and ADJ-RTI schemes \cite{wirsching2008adjoint} are here presented as two variants of RTI, with inexact sensitivity updating strategies. 
\subsection{Real-Time Iteration Scheme}
In NMPC, a NLP problem can be formulated by applying direct multiple shooting \cite{bock1984multiple} to an OCP over the prediction horizon $T=[t_0,t_f]$, which is divided into $N$ \emph{shooting intervals} $[t_0,t_1,\ldots,t_N]$, as follows
\small
\begin{subequations}\label{NLP0}
\begin{align}
\min_{s_k,u_k} \,&\sum_{k=0}^{N-1} h_k(s_k,u_k)+h_N(s_N)\\
s.t.\, &0=x_0-\hat{x}_0,\label{initial value embedding}\\
&0=x_{k+1}-\phi_k(x_k,u_k),\, k=0,1,\ldots,N-1,\label{conti const}\\
&0\geq r(x_k,u_k), \, k=0,1,\ldots,N-1,\label{path constraint}\\
&0\geq l(s_N),
\end{align}
\end{subequations}
\normalsize
where $\hat{x}_0$ is the measurement of the current state. System states $x_k\in\mathbb{R}^{n_x}$ are defined at the discrete time point $t_k$ for $k=0,\ldots,N$ and the control inputs $u_k\in \mathbb{R}^{n_u}$ are piece-wise constant. Here, \eqref{path constraint} is the inequality constraint where $r(x_k,u_k): \mathbb{R}^{n_x}\times\mathbb{R}^{n_u} \rightarrow \mathbb{R}^{n_r}$. Equation \eqref{conti const} refers to the \emph{continuity constraint} where $\phi_k(x_k,u_k)$ is a numerical integration operator that solves the following initial value problem (IVP) \footnote{For simplicity we consider Ordinary Differential Equations (ODEs) only but the extension to DAEs can be easily derived.} and return the solution at $t_{k+1}$.
\begin{equation*}
0=f(\dot{x}, x(t),u(t),t),\quad x(0)=x_k.
\end{equation*}
The NLP problem (\ref{NLP0}) depends on the state and control initialization $\mathbf{w}$ and the state measurement $\hat{x}_0$, where $\mathbf{w}=(w_0^\top,w_1^\top,\ldots,w_{N-1}^\top,x_N^\top)^\top$ and $w_k=(x_k^\top,u_k^\top)^\top$ for $k=0,\ldots,N-1$. By embedding \eqref{initial value embedding} into \eqref{conti const}, the NLP problem can be written in a compact form as 
\begin{subequations}\label{NLP}
\begin{align}
\min_{\mathbf{w}}\, &A(\mathbf{w})\\
s.t.\, & B(\mathbf{w})=0,\label{eq_G}\\
& C(\mathbf{w})\leq 0.
\end{align}
\end{subequations}

In RTI, problem (\ref{NLP}) is solved by a tailored SQP method, where only one SQP iteration is performed at each sampling instant. At sampling instant $i$, the QP subproblem initialized at $\mathbf{w}^i$ is defined as  
\begin{subequations}\label{QP0}
\begin{align}
\min_{\Delta \mathbf{w}}\, &\frac{1}{2}\Delta \mathbf{w}^\top H^i\Delta \mathbf{w}+\nabla A^i\Delta \mathbf{w}\\
s.t.\, & b^i=0,\label{update}\\
& c^i\leq 0,
\end{align}
\end{subequations}
where $\Delta \mathbf{w}=\mathbf{w}-\mathbf{w}^i$ and $\nabla$ is the gradient or Jacobian operator over $\mathbf{w}$ if no subscript is provided. The equality and inequality constraints are given by 
\begin{align*}
& b^i=B(\mathbf{w}^i)+\nabla B(\mathbf{w}^i)\Delta \mathbf{w},\\
& c^i=C(\mathbf{w}^i)+\nabla C(\mathbf{w}^i)\Delta \mathbf{w}.
\end{align*}
$H^i$ is the Hessian of the Lagrangian of \eqref{NLP}, which is defined by $\mathcal{L}(\mathbf{w},\lambda,\mu)\coloneqq A(\mathbf{w})+\lambda^\top B(\mathbf{w})+\mu^\top C(\mathbf{w})$, where $\lambda,\mu$ are Lagrangian multipliers associated with equality and inequality constraints, respectively. For most QP problems arising from NMPC, the Gauss-Newton Hessian approximation provides a sufficiently accurate Hessian with reduced computational burden \cite{diehl2002real}. Being independent of Lagrangian multipliers, the Gauss-Newton Hessian is adopted in this paper. Given the multi-stage nature of problem \eqref{NLP0}, matrices $H^i$ and $\nabla C(\mathbf{w}^i)$ are block diagonal. In particular, the Jacobian matrix $\nabla B(\mathbf{w}^i)$ has the following form:
\small
\begin{equation}\label{blockstructure}
\nabla B(\mathbf{w}^i)=\begin{bmatrix}
\mathcal{I}_{n_x}\\
\nabla \phi_0^i & -\mathcal{I}_{n_x} & \\
    & \nabla \phi_1^i & -\mathcal{I}_{n_x} & \\
    &     &   \ddots & \ddots & \\
    &     &        & \nabla \phi_{N-1}^i  & -\mathcal{I} _{n_x}
\end{bmatrix},
\end{equation}
\normalsize
where $\nabla \phi_k^i=\nabla \phi_k(x_k^i,u_k^i),\, k=0,\ldots,N-1$ and $\mathcal{I}_a$ is an identity matrix of size $a$.
The solution $(\mathbf{\Delta w}^{QP}, \lambda^{QP}, \mu^{QP})$ of \eqref{QP0} is used to update the solution of \eqref{NLP} by a single, full Newton step as
\begin{align}\label{NLP-solution-update}
\begin{split}
&\mathbf{w}^{i+1}=\mathbf{w}^{i}+\Delta \mathbf{w}^{QP},\\
&\lambda^{i+1}=\lambda^{QP}, \, \mu^{i+1}=\mu^{QP}.
\end{split}
\end{align}
Since only one QP problem is solved at each sampling instant, the RTI scheme is a special case of the linear, time-varying MPC strategy \cite{gros2016linear}.

\subsection{Inexact Sensitivity Schemes}
When formulating problem \eqref{QP0}, the Jacobian matrix $\nabla B(\mathbf{w})$ is computed at each sampling instant to obtain the current linearization of system dynamics \cite{wirsching2008adjoint}. Such computation involves sensitivity propagation of the numerical integration operator $\phi_k$ in \eqref{conti const} w.r.t. the initialization $w_k$ for each shooting interval, which can be computationally expensive for systems that are highly stiff or governed by implicit differential equations and DAEs.
To avoid the repeated sensitivity computations, in ML-RTI, firstly proposed in \cite{bock2007constrained},  the computation of $\nabla B(\mathbf{w})$ is performed at a slower rate than other components, e.g. $\nabla A$ and $B(\mathbf{w})$. Therefore, problem \eqref{QP0} with currently available but inexact sensitivities is solved at a faster rate \cite{albersmeyer2009fast, lindscheid2016parallelization}. To account for the inexact Jacobian, a so-called \emph{optimality improvement} step is employed by solving a slightly modified QP problem as
\begin{subequations}\label{QP modified}
\begin{align}
\min_{\Delta \mathbf{w}}\, &\frac{1}{2}\Delta \mathbf{w}^\top H^i\Delta \mathbf{w}+\nabla \tilde{A}^i\Delta \mathbf{w} \\
s.t.\, & \tilde{b}^i=0,\\
& c^i\leq 0,
\end{align}
\end{subequations}
where $\tilde{b}^i=B(\mathbf{w}^i)+\nabla \tilde{B}\Delta \mathbf{w}$ and $\nabla \tilde{B}$ is the Jacobian from the previous sample. The QP gradient is modified as
\small
\begin{equation*}
\nabla \tilde{A}^i= \nabla A^i+(\nabla B(\mathbf{w}^i)-\nabla \tilde{B})^\top\lambda^i,
\end{equation*}
\normalsize
where $\nabla \tilde{A}$ can be efficiently computed by applying adjoint sensitivity propagation schemes for $\nabla B^\top(\mathbf{w})\lambda$ \cite{wirsching2008adjoint, kirches2012efficient}, which are much cheaper than the computation of the full Jacobian matrix $\nabla B(\mathbf{w})$. The multi-level framework with inexact sensitivities can be summarized in Algorithm \ref{algo_inexact} \footnote{All QP components in \eqref{QP0} can be evaluated at different rates but the constraints Jacobian  only is here considered. Please refer to \cite{bock2007constrained} for the complete ML-RTI scheme.}.

\begin{algorithm}[H]	
\caption{Multi-Level inexact sensitivity RTI scheme}
\label{algo_inexact}
\begin{algorithmic}[1]
\STATE Initialize \eqref{NLP} at $(\mathbf{w}^0,\lambda^0,\mu^0)$. Choose a sensitivity update interval $m\in \mathcal{N}^+$. 
\FOR{$i=0,1,\ldots$}
    \STATE Compute $H^i,\nabla \tilde{A}^i,B^i,C^i,\nabla C^i$,
    \IF{$i \mod m=0$}
        \STATE Update the sensitivity $\nabla B^i$
        \STATE Set $\nabla \tilde{B}\leftarrow \nabla B^i$
    \ENDIF
    \STATE Solve (\ref{QP modified}) and obtain $(\mathbf{\Delta w}^{QP}, \lambda^{QP}, \mu^{QP})$
    \STATE Update the solution of the NLP problem by \eqref{NLP-solution-update}
\ENDFOR
\end{algorithmic}
\end{algorithm}
The ADJ-RTI scheme is a special variant of Algorithm \ref{algo_inexact} with $m=\infty$, in which the Jacobian matrix is computed only once off-line at the pre-defined initialization trajectory $\mathbf{w}^{0}$ \cite{wirsching2008adjoint}. When applied in the SQP framework with multiple SQP iterations, such an inexact sensitivity scheme is proved to be convergent to the local minimum of the exact sensitivity NLP problem \eqref{NLP} \cite{bock2007constrained}. The feasibility and stability of the adjoint scheme in SQP framework, without the optimality improvement, is analyzed in \cite{zanelli2016efficient}. However, there remain some open issues when applying Algorithm \ref{algo_inexact} in RTI:
\begin{itemize}
    \item It is not trivial to choose an appropriate sensitivity update interval $m$, or a pre-defined trajectory $\mathbf{w}^0$, such that the inexact Jacobian $\nabla \tilde{B}$ is a good approximation of the exact one for every sampling instant $i>0$.
    \item The sensitivities over the prediction horizon, either updated or not, are treated as a whole. Therefore, the structure of the Jacobian matrix is not exploited.
\end{itemize}
In the following Sections, a new sensitivity updating scheme is introduced that aims at overcoming the limitations described above.

\section{CMoN-RTI}\label{sec3}
Several attempts to exploit the structure of the QP \eqref{QP0} are present in the literature. A Mixed-Level scheme has been proposed in \cite{frasch2012mixed}, where only the first $N_c$ blocks in \eqref{blockstructure} are updated. However, choosing $N_c$ from heuristics may not be adequate for controlling highly nonlinear systems. Partial updating schemes where a fixed number of Jacobian blocks are updated have been independently proposed in \cite{chen2017fast} and \cite{van2017towards} by using either CMoN or a so-called ``norm criterion''. In \cite{chen2017inexact}, an inexact scheme has been proposed, where a varying number of sensitivities are updated, namely, only the most ``nonlinear'' ones. In this paper, the CMoN-RTI scheme is extended by introducing adjoint CMoN on dual variables and analyzing the QP problem by using parametric optimization theory. In particular, an advanced tuning strategy is developed, that automatically provides the number of Jacobian blocks to be updated while granting that the  DtO remains below a user-defined tolerance. For the sake of clarity, in the following the CMoN-RTI scheme proposed in \cite{chen2017inexact} is summarized.

\subsection{Curvature-like Measure of Nonlinearity}\label{CMoN sec}
Studies on Measures of the Nonlinearity (MoNs) of nonlinear dynamic systems can be traced back to the 1980s. The three main classes of MoNs are:
\begin{enumerate}
    \item the distance between a nonlinear system and its best linear approximation \cite{schweickhardt2004quantitative};
    \item the gap metric between two linear systems obtained by linearizing a nonlinear system around two different operating conditions \cite{galan2003gap};
    \item the curvature MoN (CMoN) at a point in the parameter space along a given direction. \cite{bates1980relative, guay1996measurement}.
\end{enumerate}
Global and off-line metrics are developed in \cite{schweickhardt2004quantitative, galan2003gap}. 
{CMoN is a local metric originally introduced to measure the nonlinearity in an estimation setting} \cite{bates1980relative,niu2008curvature,mallick2005differential} and then extended to chemical processes control  \cite{guay1996measurement, guay1997measurement}. It is defined as the ratio of the quadratic term over the linear term of the Taylor expansion of a nonlinear function $z=g(s)$ along the  $\epsilon$ direction in the input space:
\begin{equation}\label{original cmon}
    \kappa^o\coloneqq\frac{\norm{\ddot{z}\epsilon^2}}{\norm{\dot{z}\epsilon}^2}\:.
\end{equation}
As the scaling effect of $\epsilon$ is cancelled out by using a square norm in the denominator of $\eqref{original cmon}$, this definition evaluates the instantaneous ``curvature'' of the manifold of $z$. However, a knowledge of up to second order derivatives of the function $z$, which are computationally expensive, is required. Also, higher order terms are not taken into account \cite{li2012measure}. 

In \cite{chen2017fast, chen2017inexact}, a variant of CMoN has beeen proposed to measure the local nonlinearity of dynamic systems in the NMPC framework. Assuming that $\phi_k$ in \eqref{conti const} is twice differentiable in $w_k$, the sensitivities of $\phi_k$ w.r.t. the initialization at two consecutive sampling instants $i$ and $i-1$ satisfy
\begin{align}\label{cmon-derive}
\begin{split}
&\norm{(\nabla \phi_k^i-\nabla \phi_k^{i-1})q_k^{i-1}}\\
=&\norm{q_k^{i-1^\top}\nabla^2 \phi_k^{i-1}q_k^{i-1}+\mathcal{O}(\norm{q_k^{i-1}}^3)},\\
\approx& 2\norm{\phi_k^{i}-\phi_k^{i-1}-\nabla \phi_k^{i-1}q_k^{i-1}},
\end{split}
\end{align}
where $\norm{\cdot}$ denotes the Euclidean norm \footnote{In the paper, all vector and matrix norms are Euclidean.} and $q_k^{i-1}=w_k^i-w_k^{i-1}$ is the distance between the two initializations. The tensor $\nabla^2 \phi_k^{i-1}$ in \eqref{cmon-derive} is a vector of length $n_x$ with each element a $(n_x+n_u)$ by $(n_x+n_u)$ matrix. The computation of $q_k^{i-1^\top}\nabla^2 \phi_k^{i-1}q_k^{i-1}$ involves a vector-tensor-vector product and is defined in terms of $n_x$ vector-matrix-vector products \cite{bates1980relative}. The CMoN of $\phi_k$ is defined by
\begin{align}
\kappa_k^{i}\coloneqq\frac{\norm{\phi_k^{i}-\phi_k^{i-1}-\nabla \phi_k^{i-1}q_k^{i-1}}}{\norm{\nabla \phi_k^{i-1}q_k^{i-1}}},\label{curvature}
\end{align}
where higher order terms of $\phi_k$ are included in the numerator. Observe that,  knowledge of only the first order derivative $\nabla \phi_k^{i-1}$ is required. According to \eqref{cmon-derive}, such CMoN  measures the relative change of the directional sensitivities between two consecutive sampling instants. Observe that $\kappa_k^{i}=0$ if $\phi_k$ is linear.

Similarly, an adjoint CMoN can be defined as follows to measure the relative change of the directional sensitivities over dual variables:
\begin{align}
\tilde{\kappa}_k^{i}\coloneqq\frac{\norm{\Delta \lambda^{i-1^\top}_{k+1}(\nabla \phi_k^{i}- \nabla \phi_k^{i-1})}}{\norm{\Delta \lambda^{i-1^\top}_{k+1} \nabla \phi_k^{i-1}}}.\label{dual CMoN}
\end{align}
The term $\Delta \lambda^{i-1^\top}_{k+1}\nabla \phi_k^{i}$ can be computed by efficient adjoint sensitivity schemes. As will be shown in Sec.~\ref{sec4}, \eqref{curvature} together with \eqref{dual CMoN} play important roles in controlling the accuracy of QP solutions.

At each sampling instant, the nonlinearity of a dynamic system over the entire prediction horizon can be estimated by applying \eqref{curvature} and \eqref{dual CMoN} to each shooting interval. 

\subsection{Updating Logic}\label{sec 3 B}
Due to the multiple shooting discretization, each block $\nabla\phi_k^i$ in \eqref{blockstructure}, and the corresponding CMoN $\kappa_k^i$, uniquely depend on the initialization $w_k$. Hence,  evaluation of CMoN,  integration, and  sensitivity generation can be performed independently at each shooting interval . Thus, the set of sensitivity blocks $\{\nabla\phi_k^i\}$ can be divided into two parts: 
\begin{enumerate}
    \item an updating subset where the sensitivity blocks are updated; and
    \item the remaining subset where the sensitivity blocks are kept unchanged.
\end{enumerate}
If the first subset is much smaller than the second one, a significant reduction of computational cost for sensitivity evaluations can be achieved. To this end, CMoN can be used to determine such an updating subset. Intuitively, when $\kappa_k^{i}$ is sufficiently small, the sensitivity $\nabla \phi_k^{i}$ is close enough to $\nabla \phi_k^{i-1}$, hence sensitivity update is not necessary for the current sampling instant.  The block $k$ of the Jacobian matrix $\nabla B$ is updated according to the following strategy. Set the values of  thresholds $\eta^i_{pri}$ and $\eta_{dual}^i$, where the subscript $_{pri}$ denotes the \emph{primal variable} and $_{dual}$ the \emph{dual variable}. Then,
\begin{align}
\nabla \phi_{k}^{i}=\left\{
\begin{array}{l}
\nabla \phi_k^{i-1}, \: \text{if} \: \: \kappa_k^i\leq \eta^i_{pri}\, \&\, \tilde{\kappa}_k^{i}\leq \eta_{dual}^{i} , \\
\text{eval}(\nabla \phi_k^i),\, \text{otherwise}\
\end{array} \right. \label{Update Logic}
\end{align}

The proposed strategy is effective in both of the following cases:
\begin{enumerate}
    \item \emph{Regulation}: Given a sufficiently long prediction horizon, the system nonlinearity is typically excited in a small part of the predicted trajectory, that is, far from the steady state. As the system is approaching its steady state, less and less sensitivities are expected to be updated. 
    \item \emph{Reference tracking}: Assuming that future reference is known in advance, a widely used choice is to progressively update the reference starting from the end of the prediction horizon \cite{gros2016linear}. This approach has the beneficial impact that the initial part of the predicted trajectory is not affected by the reference change. As a result, given a sufficiently high sampling frequency, sensitivity update is necessary only in the final part of the prediction horizon, whereas information from the past can be effectively used elsewhere.
\end{enumerate}

\begin{figure*}[!t]
\begin{subequations}\label{all_matrices}
\small
\begin{align}
&M(\mathbf{0})=\begin{bmatrix}
\begin{tabular}{c:ccc:cccc}
$\nabla^2_{\Delta \mathbf{w}} \mathcal{L}_{QP}(\mathbf{0})$ & $\nabla_{\Delta \mathbf{w}} c_1^\top,$ & $\ldots,$ & $\nabla_{\Delta \mathbf{w}} c_{n_I}^\top$ & $\nabla_{\Delta \mathbf{w}} b_1^\top(\mathbf{0}),$ & $\ldots,$ & $\nabla_{\Delta \mathbf{w}} b_{n_E}^\top(\mathbf{0})$\\
\hdashline
$-\Delta \mu_1\nabla_{\Delta \mathbf{w}} c_1$   & $-c_1$ & & & \\
$\vdots$ &  &$\ddots$  & & & 0&\\
$-\Delta \mu_{n_I}\nabla_{\Delta \mathbf{w}} c_{n_I}$ &  &  &$-c_{n_I}$ & & &\\
\hdashline
$\nabla_{\Delta \mathbf{w}} b_1(\mathbf{0})$ & & & & & &\\
$\vdots$ & & 0 & & &0 &\\
$\nabla_{\Delta \mathbf{w}} b_{n_E}(\mathbf{0})$ & & & & & &\\
\end{tabular}
\end{bmatrix},\label{matrix M}\\
&N(\mathbf{0})=\begin{bmatrix}
-\nabla^2_{\mathbf{p}\Delta w}\mathcal{L}_{QP}, \Delta\mu_1\nabla_\mathbf{p} c_1^\top,\ldots, \Delta\mu_{n_I}\nabla_\mathbf{p} c_{n_I}^\top, -\nabla_\mathbf{p} b_1^\top(\mathbf{0}),\ldots, -\nabla_\mathbf{p} b_{n_E}^\top(\mathbf{0})
\end{bmatrix}^\top,\label{matrix N}\\
&\mathcal{L}_{QP}(\mathbf{0})=\frac{1}{2}\Delta \mathbf{w}^\top H\Delta \mathbf{w}+\nabla \mathcal{L}\Delta \mathbf{w}+b^\top(\mathbf{0}) \Delta \lambda+ c^\top\Delta\mu.
\end{align}
\normalsize
\end{subequations}
\hrulefill
\vspace*{4pt}
\end{figure*}

\section{An Advanced Tuning Strategy}\label{sec4}
In \eqref{Update Logic},  thresholds $\eta_{pri}$ and $\eta_{dual}$  regulate the trade-off between the accuracy of the Jacobian approximation and the computational cost, by determining the updating subset with the largest CMoN values. 
An intuitive way of choosing the thresholds is to set a constant value, i.e. $\eta^i_{pri}=\eta_{pri}^0$ and $\eta^i_{dual}=\eta_{dual}^0$, for all sampling instants.  When $\eta^0_{pri}=0$ and $\eta^0_{dual}=0$, the proposed scheme becomes the standard RTI scheme with $N_f=N$, i.e. all sensitivities are updated at every sampling instant. When $\eta^0_{pri}\geq\max(\kappa_k^i)$ and $\eta^0_{dual}\geq\max(\tilde{\kappa}_k^i)$ for all $i$, $N_f=0$ and no sensitivity is updated on-line, hence CMoN-RTI coincides with ADJ-RTI \cite{bock2007constrained}. Thresholds $\eta_{pri}^0$ and $\eta_{dual}^0$ can  take any value in the sets $[0,\max(\kappa_k^i)]$ and $[0,\max(\tilde{\kappa}_k^i)]$, respectively, to achieve a flexible tuning. 
A tuning strategy suitable for real-time implementation can be used: $\eta^i_{pri}$ and $\eta^i_{dual}$ can be chosen to update, at each instant, a fixed number of sensitivities \cite{chen2017fast, van2017towards}.
However, a pre-defined limited number of sensitivity updates may not be suitable for controlling highly nonlinear systems. 

A satisfactory trade-off between the accuracy of the sensitivity approximation and computational cost can be achieved by means of an advanced, time-varying tuning of the thresholds $\eta_{pri}$ and $\eta_{dual}$. 
The key observation is that using the inexact Jacobian in \eqref{QP modified} affects the accuracy of both primal and dual solutions. A relation that reflects inaccuracy of the sensitivities into inaccuracy of the solution of the QP problem can therefore be used to choose, at each sampling instant, the values of $\eta_{pri}$ and $\eta_{dual}$, that guarantee a tunable, bounded error on the QP solution. By adopting this strategy, CMoN-RTI can adjust the number of updated Jacobian blocks according to system operating conditions to achieve a numerical and control performance as close as possible to the standard RTI scheme, with improved computational performance.

First some facts from parametric programming theory are reviewed, then the advanced tuning strategy is detailed and some practical implementation aspects are finally considered.

\subsection{Parametric Nonlinear Programming: stability of the solution}
Two definitions concerning parametric QP are first introduced. The Jacobian approximation error is taken as a perturbation parameter. Three Lemmas describing the stability of the QP solution w.r.t. to such parameter are then given. 

\begin{mydef}\label{def1}
Define a parametric QP($\mathbf{p}$) with parameter vector $\mathbf{p}\in\mathbb{R}^{n_p}$ in the equality constraint as
\small
\begin{subequations}\label{perturbed QP}
\begin{align}
\min_{\Delta \mathbf{w}}\, &\frac{1}{2}\Delta \mathbf{w}^\top H\Delta \mathbf{w}+\nabla \mathcal{L}\Delta \mathbf{w}\\
s.t.\, & b(\mathbf{p})=0,\label{inexact constraint}\\
& c\leq 0,
\end{align}
\end{subequations}
\normalsize
where $\nabla \mathcal{L}$ is the gradient of the Lagrangian of \eqref{NLP}, $b(\mathbf{p})=B(\mathbf{w})+ (\nabla B+P)\Delta \mathbf{w}$, $P\coloneqq\nabla \tilde{B}-\nabla B$ is the Jacobian approximation error, and $\nabla \tilde{B}$ is the inexact Jacobian with partially updated blocks. The perturbation vector $\mathbf{p}=\text{vec}(P)\in \mathbb{R}^{n_p}$ is the vectorization of $P$ after eliminating zero elements.
\end{mydef}

According to Definition \ref{def1}, the exact Jacobian QP problem \eqref{QP modified} is referred to as QP($\mathbf{0}$). Due to multiple shooting discretization, $P$ has the following banded block structure
\small
\begin{equation*}
P=\begin{bmatrix}
O_{n_x}\\
P_0 &  O_{n_x} &    &    \\
    &P_1 & O_{n_x}   &    \\
    &    &\ddots& \ddots \\
    &    &    & P_{N-1} & O_{n_x}
\end{bmatrix},
\end{equation*}
\normalsize
where $O_a$ is a zero matrix of dimension $a$ and $P_k\in\mathbb{R}^{n_x\times (n_x+n_u)}$ is the $k-$th block of the Jacobian approximation, and in general is a dense matrix. 

\begin{mydef}\label{def2}
Define
\begin{equation*}
\Delta y(\mathbf{p})=(\Delta \mathbf{w}^\top(\mathbf{p}),\Delta \mu^\top(\mathbf{p}),\Delta \lambda^\top(\mathbf{p}))^\top
\end{equation*}
the solution of \eqref{perturbed QP}, where $\Delta \mathbf{w}(\mathbf{p}), \Delta \mu(\mathbf{p}),\Delta \lambda(\mathbf{p})$ are the increments of optimization variables, multipliers for inequality and equality constraints, respectively.
\end{mydef}

Observe that  QP \eqref{perturbed QP} has a modified objective gradient with respect to \eqref{QP modified}. However, it can be easily proved that these two formulations are equivalent \cite{diehl2010adjoint}. The additional computational cost can be neglected since both formulation \eqref{perturbed QP} and \eqref{QP modified} contain adjoint sensitivities in their objective. We adopt \eqref{perturbed QP} as the standard form hereafter. 

The following Lemma shows that the distance between the primal solutions of QP$(\mathbf{0})$ and QP$(\mathbf{p})$ is bounded, and the bound is of the same order of the Jacobian approximation error.
\begin{lma}\label{lma1}\cite{daniel1973stability}.
Let $\Delta \mathbf{w}(\mathbf{0})$ and $\Delta \mathbf{w}(\mathbf{p})$ minimize QP$(\mathbf{0})$ and QP$(\mathbf{p})$ over corresponding feasible sets, respectively. Then there exists constants $c$ and $\epsilon^*>0$ such that $\norm{\Delta \mathbf{w}(\mathbf{p})-\Delta \mathbf{w}(\mathbf{0})}\leq c\epsilon$ whenever $\epsilon\leq \epsilon^*$ and $\epsilon=\norm{\nabla \tilde{G}-\nabla G}=\norm{P}$.
\end{lma}

The following Lemma shows that the solution $\Delta y(\mathbf{p})$ is a unique minimizer of \eqref{perturbed QP}. Moreover, the active set is locally stable.
\begin{lma}\cite{fiacco1983introduction}\label{lma2}
Under the assumption on differentiability, second-order sufficient conditions, constraints linear independence and the strict complementary slackness condition, there exists a unique solution $\Delta y(\mathbf{p})$, which is continuously differentiable w.r.t. $\mathbf{p}$ for $\mathbf{p}$ in a neighborhood of $\mathbf{0}$. Moreover, the set of active inequality constraints is unchanged, strict complementary slackness holds, and the active constraint gradients are linearly independent at $\Delta \mathbf{w}(\mathbf{p})$.
\end{lma}

Finally, the following Lemma provides a linearly approximated relationship between the exact and inexact solutions.
\begin{lma}\cite{fiacco1983introduction}\label{lma3}
A first order approximation of $\Delta y(\mathbf{p})$ in a neighborhood of $\mathbf{p}=\mathbf{0}$ is given by 
\begin{equation*}
\Delta y(\mathbf{p})=\Delta y(\mathbf{0})+M^{-1}(\mathbf{0})N(\mathbf{0})\mathbf{p}+\mathcal{O}(\norm{\mathbf{p}}
^2)\label{first_order_appoxi}
\end{equation*}
where $M,N$ are given in \eqref{all_matrices}, and $b_k$ and $c_k$ are the $k-$th row of $b(\mathbf{p})$ and $c$, respectively.
\end{lma}

\begin{rem}
Lemma \ref{lma2} is a sufficient but not necessary condition for the results it holds. It is either not a necessary condition for Lemma \ref{lma3}. Modern studies based on perturbation theory show that the solution manifold $\Delta y(\mathbf{p})$ is nonsmooth but continuous. Therefore, $\Delta y(\mathbf{p})$ can be close enough to $\Delta y(\mathbf{0})$ even in the presence of active-set changes. The reader is referred to \cite{zavala2010real, dontchev2013euler} and references therein for more details.
\end{rem}

\subsection{First Order Error Analysis}\label{Estimating the threshold}
In the neighborhood of $\mathbf{p}=\mathbf{0}$, (\ref{first_order_appoxi}) can be rewritten as
\begin{equation}
\Delta y(\mathbf{0})=\Delta y(\mathbf{p})-M^{-1}(\mathbf{p})N(\mathbf{p})\mathbf{p}\:.\label{first_order_approximation_rewrite}
\end{equation}
As shown in Appendix A, it holds that 
\begin{equation*}
N(\mathbf{p})\mathbf{p}=\begin{bmatrix}
P^{\top}\Delta \lambda(\mathbf{p})\\
O\\
-P\Delta \mathbf{w}(\mathbf{p})
\end{bmatrix},
\end{equation*}
and, by pre-multiplying (\ref{first_order_approximation_rewrite}) by $M(\mathbf{p})$, it follows that
\begin{equation*}
M(\mathbf{p})\Delta y(\mathbf{0})=M(\mathbf{p})\Delta y(\mathbf{p})+N(\mathbf{p})\mathbf{p}.\label{premultiply}
\end{equation*}
Therefore, the DtO at the sampling instant $i$ satisfies
\begin{align}\label{DtO}
\begin{split}
\norm{e^{i}}^2\coloneqq & \norm{\Delta y(\mathbf{0}^{i})-\Delta y(\mathbf{p}^{i})}^2\\
\leq &\norm{M^{-1}(\mathbf{p}^{i})}^2\,(\norm{P^{{i}^\top}\Delta \lambda(\mathbf{p}^{i})}^2\\
&+\norm{P^{i}\Delta \mathbf{w}(\mathbf{p}^{i})}^2).
\end{split}
\end{align}
Note that, given  a finite dimensional and non-singular real matrix $M(\mathbf{p}^{i})$, its Euclidean norm
\begin{equation}\label{rho}
    \rho^{i}\coloneqq \norm{M^{-1}(\mathbf{p}^{i})}
\end{equation}
is bounded. Hence, the DtO $\norm{e^i}$ is bounded only if $\norm{P^{{i}^\top}\Delta \lambda(\mathbf{p}^{i})}$ and $\norm{P^{i}\Delta \mathbf{w}(\mathbf{p}^{i})}$ are bounded. The two bounds are referred as the \emph{dual bound} and \emph{primal bound} respectively, and are discussed in the following.

\subsubsection{Primal Bound}
By using the primal threshold $\eta_{pri}$ in the updating logic \eqref{Update Logic}, one obtains
\begin{equation}\label{jacobian error control}
\norm{P^{i} \mathbf{q}^{i-1}}\leq 2\eta_{pri}^{i} \norm{V^{i-1}_{pri}},
\end{equation}
where $\mathbf{q}^{i-1}=[q_0^{i-1^\top},\ldots,q_{N-1}^{i-1^\top}]^\top$ and $V_{pri}^{i-1}$ is a vector of directional sensitivities given by
\begin{align*}
    V_{pri}^{i-1} = [(\nabla \phi_0^{i-1}q_0^{i-1})^\top,\ldots, (\nabla \phi_{N-1}^{i-1}q_{N-1}^{i-1})^\top]^\top.
\end{align*}
Derivation details are presented in Appendix B. Moreover, there exists a $\alpha^i\geq 0\in\mathcal{R}$ such that
\begin{align}\label{alpha}
    \norm{P^{i}\Delta \mathbf{w}(\mathbf{p}^{i})} = \alpha^i\norm{P^{i} \mathbf{q}^{i-1}}
\end{align}
Hence, a bound in the direction of the primal variable is as follows
\begin{equation*}
\norm{P^{i}\Delta \mathbf{w}(\mathbf{p}^{i})}^2\leq 4\alpha^{i^2}\eta_{pri}^{i^2}\norm{V^{i-1}_{pri}}^2.
\end{equation*}

\subsubsection{Dual Bound}
Similarly, for adjoint CMoN it holds that
\begin{align*}
\norm{\Delta \lambda^\top(\mathbf{p}^{i-1})P^i}\leq \eta^{i}_{dual}\norm{V^{i}_{dual}}.
\end{align*}
where 
\begin{align*}
   V_{dual}^{i-1} = [\lambda_1^{i-1^\top} \nabla \phi_{0}^{i-1},\ldots, \lambda_{N}^{i-1^\top}\nabla \phi_{N-1}^{i-1}].
\end{align*}
There exists a $\beta^i\geq 0\in\mathcal{R}$ such that
\begin{align}\label{beta}
    \norm{P^{{i}^\top}\Delta \lambda(\mathbf{p}^{i})}\leq \beta^i\norm{\Delta \lambda^\top(\mathbf{p}^{i-1})P^i}.
\end{align}
Hence, a bound in the direction of the dual variable is obtained as follows
\begin{equation*}
\norm{P^{{i}^{\top}}\Delta \lambda(\mathbf{p}^{i})}^2\leq \beta^{{i}^2}\eta_{dual}^{i^2}\norm{V^{i-1}_{dual}}^2
\end{equation*}

\subsection{Thresholds Estimation}
Given a DtO tolerance $\bar{e}^i$ at the sampling instant $i$, let
\begin{align}\label{bound est}
\begin{split}
&\beta^{{i}^2}\eta_{dual}^{i^2}\norm{V^{i-1}_{dual}}^2\leq (1-c_1)\bar{e}^{i^2}/\rho^{i^2},\\
&4\alpha^{i^2}\eta_{pri}^{i^2}\norm{V^i_{pri}}^2\leq c_1\bar{e}^{i^2}/\rho^{i^2},
\end{split}
\end{align}
where $0<c_1<1$ is a tuning parameter that trades off impact of the primal and dual bounds on the DtO. By substituting \eqref{bound est} into \eqref{DtO}, one obtains $\norm{e^i}^2\leq \norm{\bar{e}^i}^2$. Therefore, the primal and dual thresholds satisfy the following inequalities:
\small
\begin{align}\label{eta explicit}
\begin{split}
&0\leq \eta_{pri}^{i}\leq \frac{\sqrt{c_1}\bar{e}^{i}}{2\alpha^{i}\rho^{i}\norm{V^{i-1}_{pri}}} \coloneqq \mathcal{U}_1\\
&0\leq \eta_{dual}^{i}\leq \frac{\sqrt{1-c_1}\bar{e}^{i}}{\beta^{i}\rho^{i}\norm{V^{i-1}_{dual}}} \coloneqq \mathcal{U}_2.
\end{split}
\end{align}
\normalsize
\begin{thm}\label{lm4}
$\mathcal{U}_1, \mathcal{U}_2:\mathbb{R}\times \mathbb{R}\rightarrow \mathbb{R}$ are piecewise discontinuous functions of $(\eta_{pri}^{i},\eta_{dual}^{i})$ and their ranges are finite sets.
\end{thm}
The proof of Theorem \ref{lm4} is given in Appendix C. Given Theorem \ref{lm4}, a formal solution to find the maximal $(\eta_{pri}^{i},\eta_{dual}^{i})$ is then to solve the following problem
\begin{subequations}\label{eta implicit}
\begin{align}
\max_{\eta_{pri}^{i},\eta_{dual}^{i}}\,&\eta_{pri}^{i},\eta_{dual}^{i}\\
s.t.\,& \eta_{pri}^{i}-\mathcal{U}_1(\eta_{pri}^{i},\eta_{dual}^{i}) \leq 0, \\
    & \eta_{dual}^{i}-\mathcal{U}_2(\eta_{pri}^{i},\eta_{dual}^{i}) \leq 0. 
\end{align}
\end{subequations}
The solution of problem \eqref{eta implicit} provides the maximal values of the  thresholds, corresponding to the minimum number of sensitivity updates while guaranteeing a bounded DtO. 
Note that for a given $\bar{e}^{i}\geq 0$, there always exists at least one feasible solution to \eqref{eta implicit}, i.e.  $(\eta_{pri}^{i},\eta_{dual}^{i})=0$, that makes CMoN-RTI coincide with the standard RTI scheme.

\subsection{Practical Implementation}
Problem \eqref{eta implicit} can be solved via enumeration, which requires to repeatedly solve problem \eqref{perturbed QP}. However, this is computationally prohibitive and undermining the advantage of CMoN-RTI. A practical approach to avoid solving problem \eqref{eta implicit} is setting the two thresholds at their upper bounds in \eqref{eta explicit}, by using approximated information from previous sampling instants. 

Firstly, the unknown $\rho^{i}$ in \eqref{rho} is replaced by $\rho^0$. The rationale for such choice is given by the fact that $\rho^{i}$ is the reciprocal of the smallest singular value of $M(\mathbf{p}^i)$. According to \eqref{all_matrices}, $M(\mathbf{p}^i)$ is a very sparse matrix and its smallest singular value is close to $0$ and does not vary much between sampling instants. Hence,  $\rho^0$ cab be computed offline and used for all on-line computations.

Secondly, as shown in \eqref{alpha} and \eqref{beta}, the values of $(\alpha^{i},\beta^{i})$ cannot be computed in a real time implementation, since the Jacobian approximation error $P^{i}$ cannot be computed from approximate sensitivities, and the solution $(\Delta w(\mathbf{p}^i), \Delta \lambda(\mathbf{p}^i))$ is not known in advance.  When the NMPC controller is converging on the fly, it holds that $\norm{\Delta w(\mathbf{p}^i)}\leq \norm{\mathbf{q}^{i-1}}$. In such case, $\alpha^{i}$ and $\beta^{i}$ are usually less than one. Since larger values of $\alpha^{i}$ and $\beta^{i}$ give more conservative results (as they lead to the computation of a larger number of sensitivities), a sensible choice is setting $(\alpha^{i},\beta^{i})=(1,1)$. This aspect is also discussed in Section V and VI with reference to practical implementation of the algorithm.

Thirdly, $c_1$ is the parameter that allows to balance the impact of the primal and dual thresholds on the DtO. In this paper, the choice $c_1 = 0.1$ is made since the magnitude of multipliers is typically bigger than the primal solution. As shown in Section \ref{sec5}, this choice allows to achieve a satisfactory performance. 

Fourthly, in \eqref{DtO}, an upper bound for DtO is obtained by means of norm inequality. Such inequality may lead to conservative upper bounds of thresholds \eqref{eta explicit}, hence updating more sensitivities than necessary. To account for such issue, a scaling parameter $\gamma^i$ is introduced such that
\begin{align}\label{DtO-scale}
\begin{split}
\gamma^{i^2}\norm{e^{i}}^2 =& \norm{M^{-1}(\mathbf{p}^{i})}^2\,(\norm{P^{{i}^\top}\Delta \lambda(\mathbf{p}^{i})}^2\\
&+\norm{P^{i}\Delta \mathbf{w}(\mathbf{p}^{i})}^2).
\end{split}
\end{align}
The value of $\gamma^i$ cannot be computed on-line by using \eqref{DtO-scale} for real time applications. However, an estimate of it can be obtained by relying on the following theorem.
\begin{thm}\label{thm3}
For a real, linear system $z = Xt$ where $z\in\mathcal{R}^m, t\in\mathcal{R}^m, X\in\mathcal{R}^{m\times m}$, it holds that $\norm{z}=\norm{X}\cdot\norm{t}$ if and only if $\Sigma=\sigma^2 I $,  where $X=U \Sigma V^{\top}$ is the Singular Value Decomposition (SVD) of $X$.
\end{thm}
Theorem \ref{thm3} can be proved by applying the definitions of SVD and spectral norm of matrices. According to Theorem \ref{thm3}, if the singular values of $M^{-1}(\mathbf{p}^{i})$ are all equal, \eqref{DtO-scale} holds for $\gamma^i=1$. For general matrices whose singular values are not identical, \eqref{DtO-scale} holds for $\gamma^i>1$. Hence, $\gamma^i$ is estimated by
\begin{align}\label{gamma}
     \gamma^i= \text{std}(\sigma(M^{-1}(\mathbf{p}^{i})))+1,
\end{align}
where $\text{std}(\Sigma)$ is the standard deviation operation and $\sigma(M^{-1}(\mathbf{p}^{i}))$ is the set of singular values of $M^{-1}(\mathbf{p}^{i})$. To make on-line computation feasible, $\gamma^0$ can be used, as it can be computed off-line. Effectiveness of this choice is discussed in Section \ref{sec5}.


Finally, the approximated thresholds estimates are given by
\begin{align}\label{eta practical}
\begin{split}
&\eta_{pri}^{i}=\frac{\gamma^0\sqrt{c_1}\bar{e}^{i}}{2\alpha^{i}\rho^{0}\norm{V^{i-1}_{pri}}} \\
&\eta_{dual}^{i}=\frac{\gamma^0\sqrt{1-c_1}\bar{e}^{i}}{\beta^{i}\rho^{0}\norm{V^{i-1}_{dual}}}.
\end{split}
\end{align}

\begin{figure*}[htb]
\centering
\includegraphics[width=0.75\textwidth]{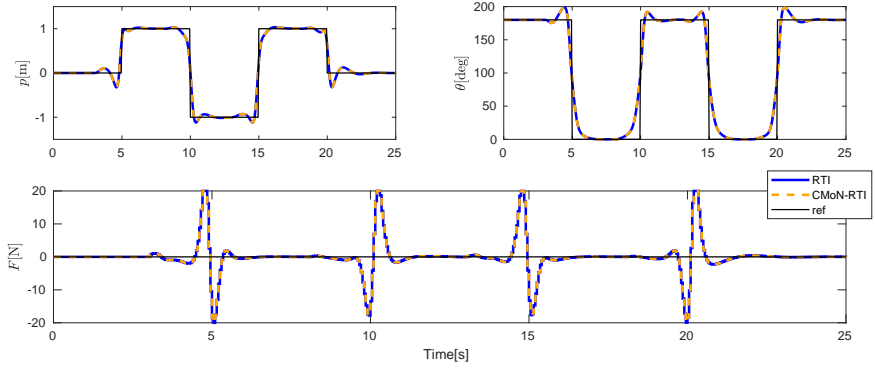}
\caption{State and control trajectories of the inverted pendulum with $N=40$. The reference signals change every 5 seconds. The constraints are $\norm{p}_\infty\leq 1$ and $\norm{F}_\infty\leq 20$. The DtO is chosen by \eqref{settings s1}. CMoN-RTI control performance is indistinguishable from that of Standard RTI. The trajectories obtained by using $N=120$ is not shown as they are identical to the ones shown in the figure.}
\label{compare_states_IP}
\end{figure*}

A summary of the practical implementation of CMoN-RTI is given in Algorithm \ref{algo dto}.
\small
\begin{algorithm}[H]	
\caption{A practical implementation of CMoN-RTI}
\label{algo dto}
\begin{algorithmic}[1]
\STATE Choose an initial point $(\mathbf{w}^0,\lambda^0,\mu^0)$
\STATE Choose $0<c_1<1$ 
\STATE Compute $\rho^0$ by \eqref{rho}
\STATE Compute $\gamma^0$ by \eqref{gamma}
\STATE Set $q_k^{-1}\leftarrow \mathbf{0}, \phi_k^{-1}\leftarrow \mathbf{0}, \nabla \phi_k^{-1}\leftarrow \mathbf{0}, w^{-1}_k\leftarrow \mathbf{0}$ for all $k$ 
\FOR{$i=0,1,\ldots$}
\STATE Compute $\nabla\mathcal{L}^i,H^i,B^i,C^i,\nabla C^i$
\FOR{$k=0,1,\ldots N-1$}
\STATE Perform integration and obtain $\phi_k^i$
\STATE Choose the DtO tolerance $\bar{e}^i$ by \eqref{tol rule}
\STATE Compute $\kappa_k^i,\tilde{\kappa}_k^i$ by \eqref{curvature} and \eqref{dual CMoN}
\STATE Update $\nabla \phi_k^i$ by \eqref{Update Logic}
\ENDFOR 
\STATE Solve QP (\ref{QP modified}) and obtain $(\Delta \mathbf{w}^{QP},\Delta \lambda^{QP},\Delta \mu^{QP})$
\STATE Update the initialization by $\mathbf{w}^{i+1}=\mathbf{w}^{i}+\Delta \mathbf{w}^{QP}, \lambda^{i+1}=\lambda^{i}+\Delta \lambda^{QP}, \mu^{i+1}=\mu^{i}+\Delta \mu^{QP}$
\STATE Compute $(\eta_{pri}^{i+1},\eta_{dual}^{i+1})$ by \eqref{eta practical}.
\ENDFOR
\end{algorithmic}
\end{algorithm}
\normalsize

\section{NMPC Simulation Case Study}\label{sec5}
In this section, Algorithm \ref{algo dto} is applied to two examples, namely, the control of an inverted pendulum and of a chain of masses. Numerical integration and sensitivity generation are performed by a $4^{\text{th}}$ order explicit Runge-Kutta integrator with $4$ steps per shooting interval, provided by the CasADi toolbox\cite{Andersson2013b} using automatic differentiation. 
The QP problem is solved by using HPIPM, a structure-exploiting interior point solver based on  hardware tailored linear algebra libraries \cite{hpipm}. Algorithmic parameters are chosen as described in Section IV.D for all examples. The computing environment is Ubuntu 16.04 on a PC with Intel core i7-4790 running at 3.60GHz, and the implementation is coded in plain C with -O2 compilation optimization flag.

\subsection{Inverted Pendulum}

An inverted pendulum is mounted on top of a cart and can roll up to 360 degrees. The dynamic model is given by
\small
\begin{align}\label{inverted pendulum}
\begin{split}
\ddot{p}&=\frac{-m_1l\sin(\theta)\dot{\theta}^2+m_1g\cos(\theta)\sin(\theta)+F}{m_2+m_1-m_1(\cos(\theta))^2},\\
\ddot{\theta}&=\frac{1}{l(m_2+m_1-m_1(\cos(\theta))^2)}(F\cos(\theta)\\
&-m_1l\cos(\theta)\sin(\theta)\dot{\theta}^2\\
&+(m_2+m_1)g\sin(\theta)),
\end{split}
\end{align}
\normalsize
where $p,\theta$ are the cart position and swinging angle, respectively, and $F$ is the control force acting on the cart. The model and values of parameters $m_1,m_2,l,g$ are taken from \cite{quirynen2015autogenerating}. For this example, a time-varying reference is given to the inverted pendulum to track different horizontal displacements and swing angles. A \emph{perfect} initialization is chosen by optimally solving the OCP for $t=0$ off-line. A short ($N=40$) and a long ($N=120$) prediction horizon are applied with a control interval $T_s=0.05$s. The tolerance on DtO in CMoN-RTI follows the rule given by 
\begin{equation}\label{tol rule}
\bar{e}^{i}=\epsilon^{abs}\sqrt{n}+\epsilon^{rel}\norm{\Delta y^i},
\end{equation}
where $(\epsilon^{abs},\epsilon^{rel})$ are the absolute and relative tolerances, $n$ is the number of optimization variables, and $y=(x,\lambda,\mu)$ is the optimal triple. Such choice ensures that the DtO tolerance scales with the size of the problem and the scale of the variable values\cite{o2013splitting}. For this problem, we set
\begin{align}\label{settings s1}
\epsilon^{abs}=10^{-1},\epsilon^{rel}=10^{-1}.
\end{align}

\begin{figure}[htb]
\centering
\includegraphics[width=0.5\textwidth]{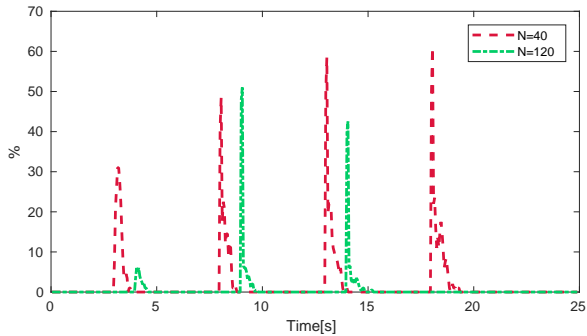}
\caption{Percentage of updated sensitivities per sampling instant. The percentage starts from $0\%$ when $N=40$ since there is no reference change within the prediction horizon in the first $3$ seconds. CMoN-RTI is able to adapt to reference changes, as can be seen from the peaks at around $t=3,8,13,18$s.}
\label{percentage}
\end{figure}

\begin{figure}[htb]
\centering
\includegraphics[width=0.5\textwidth]{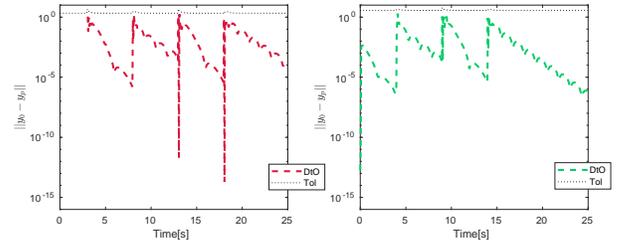}
\caption{DtO estimated on-line (colored dashed line) and the user-defined tolerance (black dotted line) for $N=40,120$, with DtO chosen in \eqref{settings s1}. The DtO increases when the system is subject to a large reference change (at around $t=3,8,13,18$s). For $N=40$, the DtO is zero in the first $3$s since there is no reference change within the prediction horizon. In all cases, the DtO is lower than the tolerance.}
\label{error_bound}
\end{figure}

\begin{figure*}[htb]
\centering
\subfloat[]{\includegraphics[height=0.2\textheight,width=.4\textwidth]{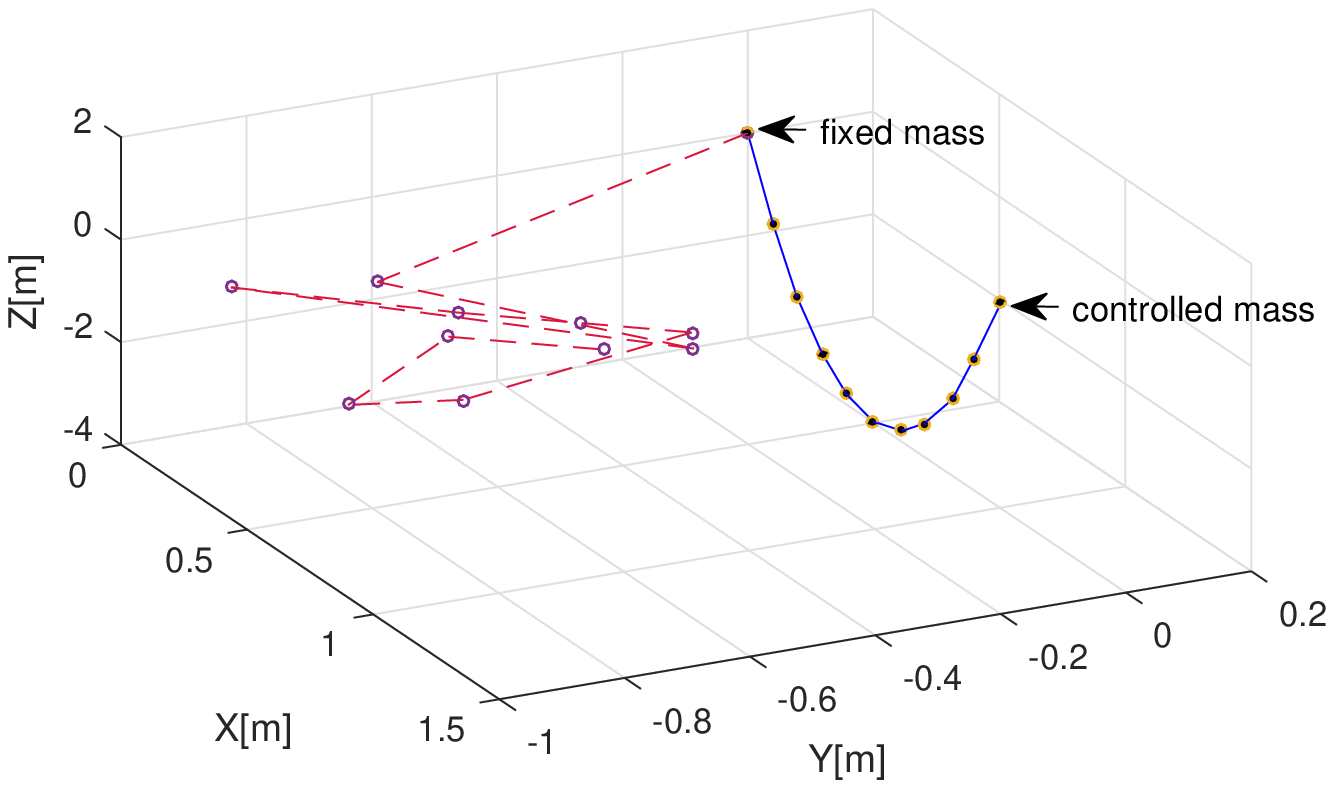}\label{fig:shape}}
\subfloat[]{\includegraphics[height=0.2\textheight,width=.6\textwidth]{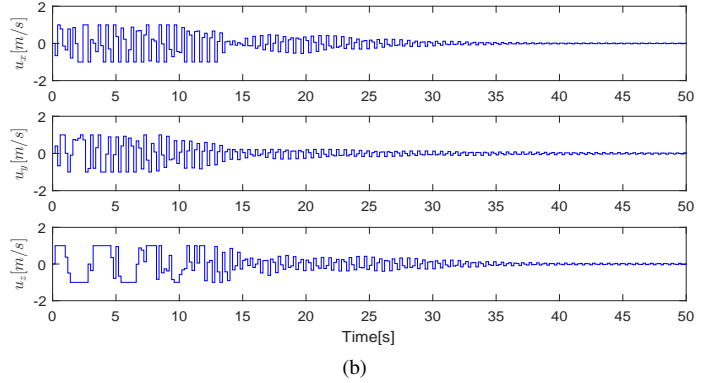}\label{fig:control}}
\caption{Initial and final positions of masses (Fig. \ref{fig:shape}) and the control trajectories (Fig. \ref{fig:control}) in one of the simulations using the standard RTI scheme. One end of the chain is fixed on a wall, while the other end is free and under control. The control interval is $T_c=0.2$s. The control inputs are constrained by $\norm{u(t)}_\infty\leq 1$.}
\label{shape and control}
\end{figure*}

\begin{figure*}[htb]
\centering
\includegraphics[height=0.2\textheight,width=1\textwidth]{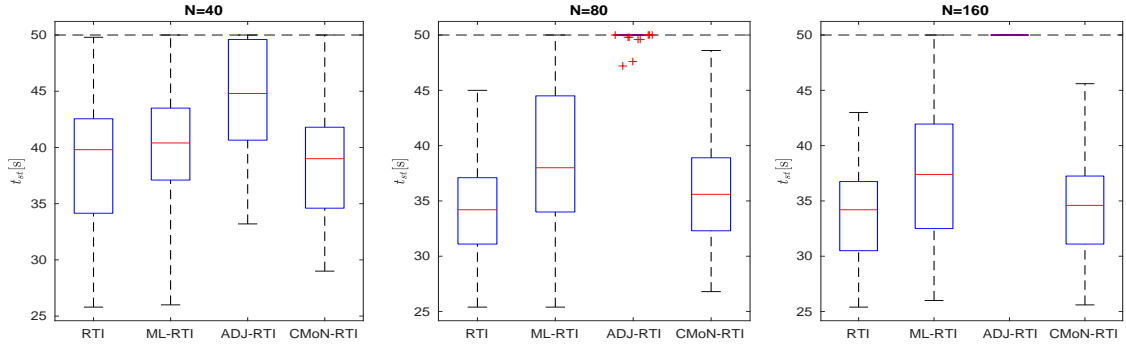}
\caption{The time $t_{st}$ needed to stabilize the chain of masses using RTI, ML- and CMoN-RTI in the total 50 simulations. The chain of masses is considered to be stabilized at time $t_{st}$ that is computed by \eqref{st}. For all schemes, the stabilizing time is set to be $t_{st}=50$s if the chain is not stabilized within $50$s.}
\label{ST_CM}
\end{figure*}

In Figure \ref{compare_states_IP}, the closed-loop state and control trajectories generated by the standard RTI and CMoN-RTI with the two prediction horizons are shown. The control performance of CMoN-RTI is indistinguishable from that of the standard RTI scheme, which demonstrates that CMoN-RTI is able to maintain the closed-loop performance as the standard RTI while using much less sensitivity computations.

In Figure \ref{percentage}, the percentage of exactly computed sensitivities per sampling instant is given. The CMoN-RTI scheme can adapt to operating conditions by evaluating more sensitivities when the reference is about to change, as the peaks occur at around $t=3,8,13,18$s. A significant reduction of the percentage of updated sensitivities is observed when $N=120$, making CMoN-RTI adequate to deal with the case of long prediction horizons. 
As explained in Section \ref{sec 3 B} , only the last part of the reference is triggering sensitivity updates. Hence, the longer the prediction horizon, the lower the percentage of sensitivities to be updated.
Figure \ref{error_bound} shows the DtO at each sampling instant together with the user-defined tolerance. An additional QP with exact Jacobian matrix is solved at each sampling instant to compute the DtO. In both cases, the DtO is lower than the tolerance. 

To examine the effectiveness of using $\rho^0$ and $\gamma^0$ in \eqref{eta practical}, the relative difference is defined as
\begin{align*}
    r\coloneqq \frac{|\frac{\gamma^i}{\rho^i}-\frac{\gamma^0}{\rho^0}|}{\frac{\gamma^i}{\rho^i}}
\end{align*}
For the inverted pendulum example, it is observed that the maximum value of $r$ is $24\%$, i.e. a sufficiently small difference that confirms the effectiveness of the approximation strategies discussed in Section \ref{sec4}.

\subsection{Chain of Masses with Nonlinear Springs}

\begin{figure*}[htb]
\centering
\includegraphics[height=0.2\textheight,width=1\textwidth]{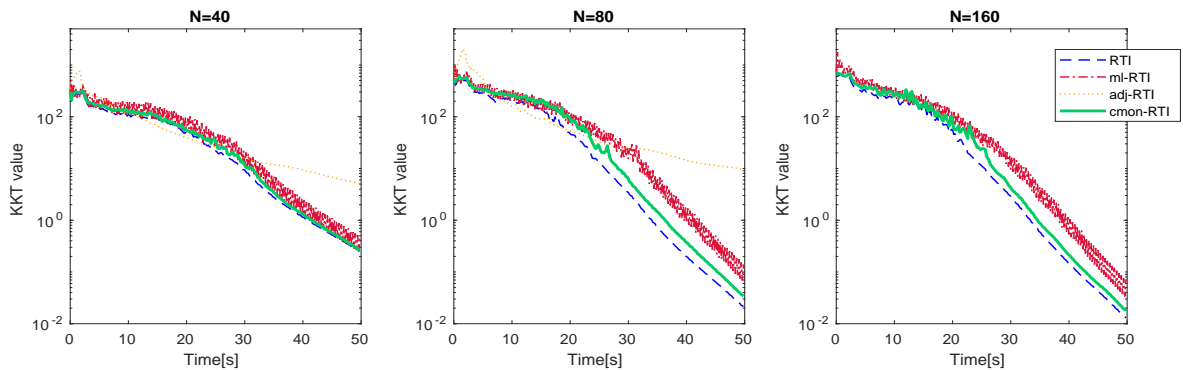}
\caption{The average KKT value at every sampling instant of the successfully stabilized cases among 50 simulations using RTI, ML- and CMoN-RTI. The KKT value is computed as the norm of the Lagrangian of the NLP \eqref{NLP} as an indicator of optimality.}
\label{KKT_CM}
\end{figure*}

A chain of masses is a system with $n$ masses connected by springs on a chain \cite{wirsching2006fast}. The dynamic model is given by
\small
\begin{align*}
&\dot{p}_i(t)=v_i(t), \, i=1,\ldots, n-1,\\
&\dot{v}_i(t)=\frac{1}{m}(F_{i+1}(t)-F_i(t))-g,\\
&\dot{p}_n(t)=u(t),
\end{align*}
\normalsize
where $p_i(t)\in\mathbb{R}^3$ and $v_i(t)\in\mathbb{R}^3$ are the positions and velocities of the $i-$th mass, respectively, and 
\small
\begin{equation*}
F_i(t)=D(x_i(t)-x_{i-1}(t))(1-\frac{L}{\norm{x_i(t)-x_{i-1}(t)}_2})+F_{NL}\;,
\end{equation*}
\normalsize
is the spring force from mass $i$ to $i+1$ and $F_{NL}$ is its nonlinear component. The velocities of the free mass $\dot{p}_n(t)$ are controlled by $u(t)$. As demonstrated in \cite{wirsching2006fast, kirches2012efficient}, ADJ-RTI is able to stabilize the chain of masses if $F_{NL}=0$, i.e. the chain is connected by linear springs. In this paper,  nonlinear springs \cite{enns2010s} are considered with  
\small
\begin{align*}
F_{NL}=D_1(x_i(t)-x_{i-1}(t))\frac{(\norm{x_i(t)-x_{i-1}(t)}_2-L)^3}{\norm{x_i(t)-x_{i-1}(t)}_2}.
\end{align*}
\normalsize
A total of $50$ simulations are performed while using the standard, ML-, ADJ- and CMoN-RTI, with randomly assigned initial positions and velocities of the masses, see e.g. Fig \ref{shape and control}, for the positions and control trajectories generated in one of the simulations. For ML-RTI, the entire constraint Jacobian matrix is updated every $m=2$ sampling instants; For ADJ-RTI, the Jacobian matrix is computed off-line at the steady state trajectory; For CMoN-RTI, the DtO tolerance is chosen as in \eqref{settings s1}. To ensure that an accurate representation of the system is always used in the controller, at least $10\%$ sensitivities are updated at each sampling instant. These sensitivities are those having the largest values of CMoN, hence exhibiting the most significant nonlinearities \cite{chen2017fast}. 

\begin{table*}[!t]
\centering
\caption{The avarage and maximal computational time per sampling instant in milliseconds[ms] for CMoN-RTI and the standard RTI scheme for the chain of masses with prediction length $N=40,80,160$. \emph{Sens.} stands for sensitivity evaluation time and \emph{QP.} is the QP solving time.}
\label{CPT}
\begin{tabular}{c|cccccccccccc|l}
\hline
\multirow{3}{*}{N} & \multicolumn{6}{c}{Average} & \multicolumn{6}{c|}{Maximal} & \multirow{2}{*}{Speedup factor} \\
 & \multicolumn{3}{c}{CMoN-RTI} & \multicolumn{3}{c}{RTI} & \multicolumn{3}{c}{CMoN-RTI} & \multicolumn{3}{c|}{RTI} &  \\
 & Total & Sens. & QP. & Total & Sens. & QP. & Total & Sens. & QP. & Total. & Sens. & QP. &  \\
\cmidrule(l){1-1} \cmidrule(l){2-4}\cmidrule(l){5-7}\cmidrule(l){8-10}\cmidrule(l){11-13}\cmidrule(l){14-14}
40 & 17.9 & 7.5 & \multicolumn{1}{c|}{8.8} & 23.5 & 14.1 & \multicolumn{1}{c|}{7.9} & 37.1 & 19.0 & \multicolumn{1}{c|}{16.1} & 40.6 & 24.2 & 14.7 & 9.4\% \\
80 & 29.7 & 9.6 & \multicolumn{1}{c|}{17.8} & 43.5 & 24.4 & \multicolumn{1}{c|}{16.2} & 61.4 & 25.5 & \multicolumn{1}{c|}{34.4} & 70.1 & 37.4 & 31.3 & 14.2\% \\
160 & 66.8 & 13.8 & \multicolumn{1}{c|}{46.3} & 93.8 & 45.4 & \multicolumn{1}{c|}{43.3} & 134.7 & 30.8 & \multicolumn{1}{c|}{98.7} & 144.8 & 52.4 & 94.0 & 7.5\% \\ \hline
\end{tabular}
\end{table*}

\begin{figure}[!htb]
\centering
\includegraphics[width=0.5\textwidth]{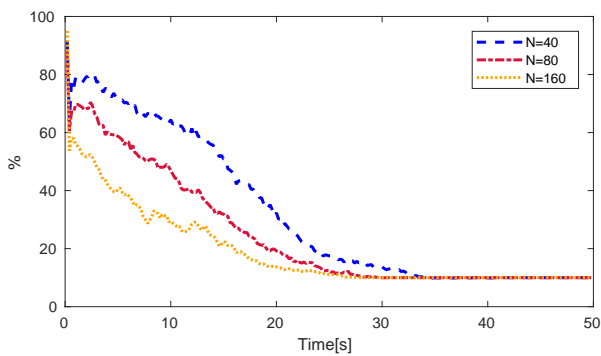}
\caption{The average percentage of exactly updated sensitivities at every sampling instant of the stabilized cases among 50 simulations for chain of masses using CMoN-RTI. At least $10\%$ of sensitivities are updated at each sampling instant.}
\label{Percentage_CM}
\end{figure}

Control performance, numerical robustness, and efficiency of CMoN-RTI are evaluated and compared with standard RTI, ML-RTI, and ADJ-RTI. Firstly we collect statistics of the \emph{stabilizing time} $t_{st}$, defined as 
\begin{subequations}\label{st}
\begin{align}
& t_{st}=\argmin t\\
s.t. & \norm{u(t_i)}_\infty<0.1, \forall t_i\geq t,
\end{align}
\end{subequations}
from the 50 simulations. In Fig. \ref{ST_CM}, the statistics of the standard, ML-, ADJ- and CMoN-RTI with $N=40,80,160$ are shown. Note that, if the chain is not stabilized within $50$s, we set $t_{st}=50$s, which is a conservative choice since the stabilization process may take far more than $50$s.  For all simulations, RTI is able to stabilize the chain within $50$s. The mean and interquartile range (IQR) of $t_{st}$ of CMoN-RTI is very close to those of the standard RTI. This means that CMoN-RTI has a similar control performance to the standard RTI in most of the situations. On the other hand, ML-RTI has a similar stabilizing time to RTI when $N$ is short, whereas $t_{st}$ grows significantly as $N$ becomes larger. ADJ-RTI, initialized at the steady state trajectory, is not able to provide acceptable control performance, especially when $N$ is large. 

The control performance is also evaluated by assessing the optimality of each controller. In Fig. \ref{KKT_CM}, the average Karush-Kuhn-Tucker (KKT) value, i.e. the norm of the gradient of the Lagrangian of the NLP \eqref{NLP}, at each sampling instant is presented. It can be observed that
\begin{itemize}
    \item ML-RTI KKT values exhibit strongly oscillatory behavior since the Jacobian update is performed every $m=2$ sampling instants only.
    \item As the system converges ``on the fly'', the KKT of CMoN-RTI decreases smoothly as that of the standard RTI.
\end{itemize}

\begin{table}[htb]
\centering
\caption{The number of simulations (among 50) where each controller \emph{cannot} stabilize the chain.}
\label{No. fail}
\normalsize
\begin{tabular}{c|ccc}
\hline
\backslashbox{}{N}  & 40 & 80 & 160 \\ \hline
ML-RTI   & 4  & 5  & 7   \\
ADJ-RTI  & 7  & 42 & 50  \\
CMoN-RTI & 0  & 0  & 0   \\ \hline
\end{tabular}%
\end{table}

As for  numerical robustness, the number of simulations where each controller fails to stabilize the system within $50$s is reported in Table \ref{No. fail} . Given that the initial condition of each simulation is randomly assigned, the numerical robustness or the sensitivity w.r.t. initialization of each controller can be assessed. CMoN-RTI is able to stabilize the chain within $50$s in all situations, although the maximal stabilizing time is larger than that of the standard RTI (see Fig. \ref{ST_CM}). ML-RTI has a few failed cases if $N=40$ and this number increases as the prediction horizon grows. Not surprisingly, ADJ-RTI exhibits the poorest robustness properties as it heavily depends on the quality of the off-line Jacobian matrix.

To evaluate computational efficiency, the average percentage of exactly updated sensitivities using CMoN-RTI at every sampling instant is reported in Figure \ref{Percentage_CM}. In the first 20s, the KKT values of CMoN-RTI and RTI are almost identical, however, the number of updated sensitivities is at most $80\%$ and it reduces to $60\%$ when $N$ becomes larger. After $t=30$s, when the system is close to its steady state, updating only $10\%$ of the blocks allows to still maintain small KKT values. Table \ref{CPT} shows the average and maximal computational time of CMoN-RTI and the RTI scheme per sampling instant.
For this example, the speedup factor is computed by using the maximal computational time. With different prediction horizons, the computational time for evaluating sensitivities varies from about $60\%$ to $36\%$ of the full RTI step, and the one for the QP varies from about $36\%$ to $65\%$. As a result, the speed up factor is strongly related to the distribution of computational time among the critical steps of the full RTI. In the specific example, the speedup factor is always positive with a maximum value of $14.2\%$ when $N=80$. Also, observe that the computational performance obtained in the examples is related to the use of the simple explicit Runge-Kutta integrator. For systems that require the use of more complex integrators, whose sensitivities are more computational expensive, CMoN-RTI is expected to achieve a greater speedup factor. 


\section{Convergence Analysis}\label{sec6}
Algorithm \ref{algo dto} is a partial sensitivity updating scheme in the framework of RTI between two consecutive sampling instants. It can also be straightforwardly extended to the SQP framework, where a sequence of QP problems is solved until convergence is achieved. The resulting algorithm, denoted hereafter as CMoN-SQP, partially updates sensitivities between two consecutive SQP iterations. In the SQP scenario, the Two-side-Rank-One (TR1) updating SQP algorithm has been proposed in \cite{griewank2002constrained} for equality constrained problems. Similar to the famous Symmetric-Rank-One (SR1) updating scheme \cite{nocedal2006numerical}, the TR1 scheme requires Hessian and Jacobian updates to satisfy both direct and adjoint secant conditions. This method is extended to linearly inequality constrained problems in \cite{diehl2010adjoint} and its local convergence is proved. 

Differently from the TR1 scheme, which adopts a rank one Jacobian matrix update, CMoN-SQP achieves a block update by exploiting the structure of the problem. In addition, the primal and dual bounds are satisfied, instead of enforcing secant conditions. In the following, local convergence of CMoN-SQP is proved and it is shown that the convergence rate is tunable via the choice of the DtO tolerance.

\subsection{Local Convergence of CMoN-SQP}
Consider the parametric QP problem \eqref{perturbed QP}. Solving problem \eqref{perturbed QP} in a SQP algorithm is equivalent to solving the following nonlinear system:
\begin{equation*}
F(\mathbf{y})=0, \mathbf{y}\coloneqq\begin{bmatrix} \mathbf{w}\\\lambda \end{bmatrix}, F(y)=\begin{bmatrix}
R^\top\nabla \mathcal{L}(\mathbf{w},\lambda)\\
B(\mathbf{w})\\
C_a(\mathbf{w})
\end{bmatrix},
\end{equation*}
where $\lambda$ denotes the multiplier for both equality and active inequality constraints, $C_a$ contains the active constraints, and $R$ is a matrix with orthonormal column vectors, such that $\nabla C_aR=0$ \cite{diehl2010adjoint}. The Jacobian matrix of the nonlinear system is
\begin{equation*}
\nabla F(\mathbf{y}^i)=\frac{\partial F}{\partial \mathbf{y}}(\mathbf{y}^i)=\begin{bmatrix}
R_i^\top H^i & R_i^\top\nabla B^\top(\mathbf{w}^i)\\
\nabla B(\mathbf{w}^i) & \\
\nabla C_a(\mathbf{w}^i) & 
\end{bmatrix},
\end{equation*}
where $H^i$ is an approximation of the exact Hessian, i.e. the Gauss-Newton approximation which is independent of the multiplier $\lambda$. Let $J_i$ be an approximation of the exact Jacobian $\nabla F(\mathbf{y}^i)$ with 
\begin{equation*}
J_i=\begin{bmatrix}
R_i^\top H^i & R_i^\top\nabla B^\top(\mathbf{w}^i)\\
\nabla \tilde{B}(\mathbf{w}^i) & \\
\nabla C_a(\mathbf{w}^i) & 
\end{bmatrix}.
\end{equation*}
The following theorem indicates that the proposed scheme is convergent in the neighborhood of $\mathbf{p}=\mathbf{0}$. 

\begin{thm}\label{thm2}
Let $F:\mathcal{V}\rightarrow \mathbb{R}^{n_y}, \mathcal{V}\subset \mathbb{R}^{n_y}$ be continuously differentiable. Consider the two sequences
\begin{align*}
&\{y_*\}:\, y_*^{i+1}=y_*^{i}+\Delta y_*^i\\
&\{y_p\}:\, y_p^{i+1}=y_p^{i}+\Delta y_p^i
\end{align*}
where
\begin{align}
&\Delta y_*^i=-\nabla F^{-1}(y_*^{i})F(y_*^{i})\label{y_*}\\
&\Delta y_p^i=-J^{-1}(y_p^{i})F(y_p^{i})\nonumber
\end{align} 
Assume that
\begin{enumerate}
    \item the Jacobian matrix is invertible, uniformly bounded, and has uniformly bounded inverses,
    \item there exists a $\kappa_0<1$ such that $\norm{\Delta y_*^{i+1}}\leq \kappa_0 \norm{\Delta y_*^{i}}$ for all $i>m_1, m_1\in\mathbb{N}$. Hence, starting from $y^0\in \mathcal{V}$, the sequence $\{y_*\}$ converges to a local optimizer $y_*^+$,  \label{assump 3}
    \item $J(y_p^i)$ is generated by Algorithm \ref{algo dto},
\end{enumerate}
Then,
\begin{enumerate}
    \item there always exists a set of scalars $\{i\in \mathbb{N}^+|\bar{e}^{i}\geq 0\}$ such that the distance between the sequences $\{y_p\}$ and $\{y_*\}$ is sufficiently small at each iteration, \label{arg1}
    \item there always exists a set of scalars $\{i\in \mathbb{N}^+|\bar{e}^{i}\geq 0\}$ and a $\kappa_2$ satisfying $\kappa_0\leq \kappa_2<1$, such that $\norm{\Delta y_p^{i+1}}\leq \kappa_2 \norm{\Delta y_p^{i}}$ for all $i>m_2, m_2\in \mathbb{N}$, and the sequence $\{y_p\}$ converges to $y_p^+=y_*^+$ starting from $y^0$.
\end{enumerate}
\end{thm}

\begin{IEEEproof}
Let the locally \emph{exact} solution initialized at $y_p^i$ be
\begin{equation}\label{y_0}
\Delta y_0^i=-\nabla F^{-1}(y_p^{i})F(y_p^{i})\:.
\end{equation}
Assume that at  iteration $i$, the DtO is satisfied as
\begin{equation*}
\norm{\Delta y_p^i-\Delta y_0^i}=\norm{e^i}\leq \bar{e}^i\:.
\end{equation*}
Let $\norm{d_y^i}=\norm{y_p^i-y_*^i}$ be the distance between the two sequences at the current iteration. Observe that
\begin{align*}
&\nabla F(y_p^i)=\nabla F(y_*^i)+d_y^{i^\top}\nabla^2 F(y_*^i)+\mathcal{O}(\norm{d_y^i}^2),\\
&F(y_p^i)=F(y_*^i)+\nabla F(y_*^i)d_y^i+\mathcal{O}(\norm{d_y^i}^2).
\end{align*}
Assume that $\norm{d_y^i}$ is sufficiently small and $\mathcal{O}(\norm{d_y^i}^2)$ can be neglected, then by combining \eqref{y_*} and \eqref{y_0}, it follows that
\begin{align*}
\Delta y_0^i-\Delta y_*^i=-\nabla F^{-1}(y_*^i)(d_y^{i^\top}\nabla^2 F(y_*^i)\Delta y_0^i)-d_y^i\:.
\end{align*}
As a result, 
\begin{equation*}
\norm{\Delta y_0^i-\Delta y_*^i}\leq g^i\norm{d_y^i}\:,
\end{equation*}
where $g^i=\norm{\nabla F^{-1}(y_*^i)(\Delta y_0^{i^\top}\nabla^2 F(y_*^i))+\mathcal{I}}$. The distance between the two solutions at the current iteration is
\begin{align*}
\norm{\Delta y_p^i-\Delta y_*^i}&\leq \norm{\Delta y_p^i-\Delta y_0^i}+\norm{\Delta y_0^i-\Delta y_*^i}\\
&\leq \underbrace{\bar{e}^i+g^i\norm{d_y^i}}_{\norm{d_{\Delta y}^i}}\:,
\end{align*}
and the distance between the two sequences at the next iteration is
\begin{align*}
\norm{d_y^{i+1}}&\coloneqq \norm{y_p^{i+1}-y_*^{i+1}}\\
&\leq \norm{d_y^i}+\norm{\Delta y_p^i-\Delta y_*^i}\\
&\leq \bar{e}^i+(1+g^i)\norm{d_y^i}.
\end{align*}
Since Algorithm \ref{algo dto} always starts from $\norm{d_y^0}=0$, $\norm{d_y^i},\forall i>0$ is a linear combination of $(\bar{e}^0,\bar{e}^1,\ldots,\bar{e}^i)$. Therefore, it is always possible to choose a set of scalars $\{i\in \mathbb{N}^+|\bar{e}^{i}\geq 0\}$, such that $\norm{d_y^i}\approx 0$. Equivalently, the sequence $\{y_p\}$ can be sufficiently close to $\{y_*\}$ at every iteration.

Consider now the convergence properties of $\{y_p\}$. By assumption \ref{assump 3}, it follows that
\begin{align*}
\norm{y_p^{i+1}}&\leq \norm{d_{\Delta y}^{i+1}}+\kappa_0\norm{\Delta y_*^i}\\
&\leq \norm{d_{\Delta y}^{i+1}}+\kappa_0 \norm{d_{\Delta y}^{i}}+\kappa_0\norm{\Delta y_p^i}\\
&=\kappa_1+\kappa_0\norm{\Delta y_p^i}
\end{align*}
where $\kappa_1=\norm{d_{\Delta y}^{i+1}}+\kappa_0 \norm{d_{\Delta y}^{i}}$. Since $\kappa_0<1$ and $\norm{d_{\Delta y}^{i+1}}, \norm{d_{\Delta y}^{i}}$ can be arbitrarily small, there exists a $\kappa_2$ satisfying $\kappa_0\leq \kappa_2<1$ such that
\begin{equation*}
\norm{y_p^{i+1}}\leq \kappa_1+\kappa_0\norm{\Delta y_p^i}\leq \kappa_2\norm{\Delta y_p^i} \:.
\end{equation*}
Therefore, the sequence $\{y_p\}$ is convergent and its convergence rate is at most identical to that of $\{y_*\}$. As proved in \cite{bock2007constrained, diehl2010adjoint}, when $\{y_p\}$ does converge, it converges to the exact limit $y_*^+$ of the sequence $\{y_*\}$.
\end{IEEEproof}

Theorem \ref{thm2} shows that the Jacobian approximation error can be controlled by using user-defined DtO tolerances, hence the convergence can be satisfied by using appropriate tuning configurations. The convergence rate is also shown to be tunable, which increases the flexibility of the proposed algorithm. If $\bar{e}^{i}=0,\forall i\geq 0$, CMoN-SQP becomes the standard SQP algorithm with the same convergence rate. 

\subsection{Numerical Examples}
As an example, the CMoN-SQP scheme is applied to the inverted pendulum \eqref{inverted pendulum}. The control objective is to invert the pendulum from bottom to top. CMoN-SQP is used to solve the OCP in open-loop at time $t=0$ with $N=40$. Since only local convergence is of interest, the initialization of the OCP is in a neighborhood of the optimal solution and a full Newton-step is adopted at each iteration. 

\begin{figure}[htb]
\centering
\includegraphics[width=.5\textwidth]{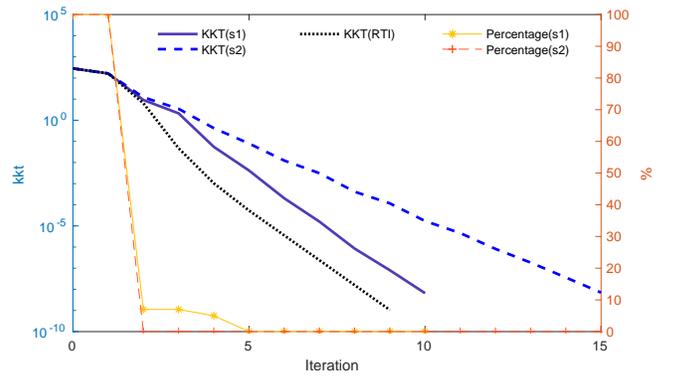}\label{fig:cmon_rti_convergence}
\caption{Convergence behavior of CMoN-SQP when applied to the inverted pendulum using $c_1=0.1$ and two DtO tolerances. The left y-axis reports the KKT value that indicates the optimality of the solution. The right y-axis reports the percentage of sensitivities being updated at each iteration. }
\label{convergence}
\end{figure}

\begin{figure}[htb]
\centering
\includegraphics[width=.5\textwidth]{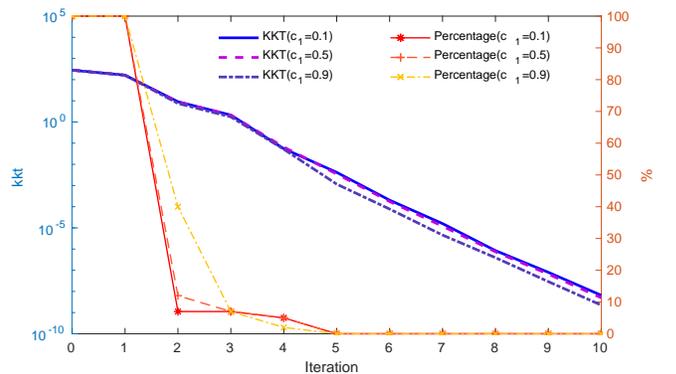}\label{fig:cmon_rti_convergence_c1}
\caption{Convergence behavior of CMoN-SQP when applied to the inverted pendulum using $(s_1)$ for DtO and three values of $c_1$. The left y-axis reports the KKT value that indicates the optimality of the solution. The right y-axis reports the percentage of sensitivities being updated at each iteration.}
\label{convergence_c1}
\end{figure}

Figure \ref{convergence} shows the convergence behavior of two different DtO choices of CMoN-SQP. The left y-axis reports the KKT value, that indicates the optimality of the solution. The right y-axis reports the percentage of sensitivities being updated at each iteration. To show how the choice of DtO tolerance affect the convergence rate, the following  DtO tolerances are used:
\begin{align*}
&s_1:\,(\epsilon^{abs}=10^{-2},\epsilon^{rel}=10^{-2})\\
&s_2:\, (\epsilon^{abs}=10^{-1},\epsilon^{rel}=10^{-1})
\end{align*}
A more aggressive setting ($s_2$) leads to less sensitivity evaluations but slower convergence rate. Figure \ref{convergence_c1} shows the convergence behavior of three choices of $c_1$ for \eqref{eta practical}. The convergence rate is not sensitive to the values of $c_1$. In practice, to solve a structured NLP problem by using CMoN-SQP, one would achieve a satisfactory trade-off between the cost of sensitivities and the number of iterations by properly tuning the DtO tolerance.


\section{Conclusion}\label{sec7}
In this paper, the partial sensitivity updating scheme CMoN-RTI of \cite{chen2017inexact} is extended by proposing an advanced tuning strategy with solution accuracy control and convergence analysis. In CMoN-RTI, sensitivities are updated based on CMoN of the dynamic system over the prediction horizon. The CMoN works as a metric to evaluate the quality of sensitivity approximation, and an updating logic is designed with the use of primal and dual thresholds. By using parametric optimization theory, an advanced strategy for tuning such thresholds is proposed. Such strategy automatically determines the number of updated sensitivities and guarantees the DtO of QP solutions under a user-defined tolerance. 

Closed-loop simulations show that the CMoN-RTI scheme exhibits good control performance when applied to highly nonlinear control problems. The application to an inverted pendulum shows that CMoN-RTI can adapt to reference changes while satisfying the DtO tolerance. The results from a chain of masses with nonlinear springs demonstrate the superior control performance, numerical robustness, and efficiency of CMoN-RTI.

The proposed scheme has also been extended to full SQP algorithms, denoted as CMoN-SQP and its local convergence is proved. Comparing to existing inexact sensitivity SQP methods, CMoN-SQP has two unique properties, namely, tunable convergence rate and structure exploiting updating logic. 

Future studies may focus on possible extensions of CMoN-RTI. While ADJ-RTI and Mixed-Level schemes can benefit from condensing steps with significantly reduced computational efforts \cite{kirches2012efficient, frasch2012mixed}, CMoN-RTI usually requires to perform a full condensing step at every sampling instant. Further improvements can be achieved by adopting  partial condensing methods \cite{kouzoupis2015block}.

\appendices
\section{}
In this Appendix, the computation of $M,N$ in (\ref{matrix M}) and (\ref{matrix N}) is detailed. For elements in $M$, it holds
\footnotesize
\begin{align*}
&\nabla^2 \mathcal{L}_{QP}=H,\\
&\nabla c_k=\nabla C_k, k=1,\ldots, n_I,\\
&\nabla b_j(\mathbf{p})=\nabla B_j+P_{j,:}, j=1,\ldots, n_E,
\end{align*}
\normalsize
where $P_{j,:}$ is the $j$th row of $P$. For elements in $N$, it holds
\footnotesize
\begin{align*}
&\nabla^2_{\mathbf{p}\Delta w}\mathcal{L}=\begin{bmatrix}
O_1 & &\\
\vdots\\
O_1 & &\\
 \Lambda_{n_x+1} &\\
 \vdots &\\
 \Lambda_{2n_x} &\\
 &  & \ddots \\
 &  &  & \Lambda_{Nn_x+1} & O_2\\
 &  &  & \vdots & \vdots\\
 &  &  & \Lambda_{(N+1)n_x}& O_2\\
\end{bmatrix},\\
&\nabla_{\mathbf{p}}c_j=O_3\in\mathbb{R}^{1\times n_p},j=0,\ldots, n_I,\\
&\nabla_{\mathbf{p}}b=\blkdiag{-\mathcal{W}_0,\ldots, -\mathcal{W}_{N-1}},
\end{align*}
\normalsize
where \footnotesize $O_1\in\mathbb{R}^{(n_x+n_u)\times (n_x+n_u)}, O_2\in\mathbb{R}^{(n_x+n_u)\times n_x}, \Lambda_j=I_{n_x+n_u}\otimes \Delta \lambda_j$ \normalsize and \footnotesize
$\mathcal{W}_k=I_{n_x}\otimes \Delta w_k^\top$ with $\Delta w_k^\top\in\mathbb{R}^{1\times (n_x+n_u)}$.

\section{}

\normalsize
In this Appendix, the expression for $V^{i-1}_{pri}$ in \eqref{jacobian error control} is derived. From the updating logic \eqref{Update Logic}, it can be easily obtained that
\footnotesize
\begin{align*}
\norm{P_k^{i}q_k^{i-1}}\leq 2\eta^{i}_{pri} \norm{\nabla \phi_k^{i-1}q_k^{i-1}}.
\end{align*}
\normalsize
For the full Jacobian matrix, it holds
\footnotesize
\begin{align*}
\norm{P^{i} \mathbf{q}^{i-1}}
&=\bignorm{\begin{bmatrix}
O\\
P_0^{i} & O   &    &    \\
    &P_1^{i} &    &    \\
    &    &\ddots & O &    \\
    &    &    & P_{N-1}^{i} & O
\end{bmatrix}\begin{bmatrix}
q_0^{i-1}\\
q_1^{i-1}\\
\vdots\\
q_{N-1}^{i-1}\\
q_N^{i-1}
\end{bmatrix}}\\
&=\bignorm{\begin{bmatrix}
O\\
P_0^{i}q_0^{i-1}\\
\vdots\\
P_{N-1}^{i}q_{N-1}^{i-1}
\end{bmatrix}}=\sqrt{\sum_{k=0}^{N-1}\norm{P_k^{i}q_k^{i-1}}^2}\\
&\leq  \sqrt{\sum_{k=0}^{N-1}4 (\eta^i_{pri})^2 \norm{\nabla \phi_k^{i-1}q_k^{i-1}}^2 }\\
&=2\eta^i_{pri} \sqrt{\sum_{k=0}^{N-1}\norm{\nabla \phi_k^{i-1}q_k^{i-1}}^2 }\\
&=2\eta^i_{pri}\bignorm{\begin{bmatrix}
\nabla \phi_0^{i-1}q_0^{i-1}\\
\nabla \phi_1^{i-1}q_1^{i-1}\\
\vdots\\
\nabla \phi_k^{i-1}q_{N-1}^{i-1}\\
\end{bmatrix}}\\
&=2\eta^i_{pri} \norm{V^{i-1}_{pri}}.
\end{align*}
\normalsize
A similar derivation can be conducted for $V_{dual}^{i-1}$. The details are hence omitted. 

\section{}
A proof of Theorem \ref{lm4} is given in this Appendix.
\begin{IEEEproof}
According to \eqref{Update Logic}, given the primal and dual thresholds, $\nabla \phi_j^i$ are updated for
\begin{align*}
    j\in\{j|\kappa_j^i>\eta_{pri}^i, \tilde{\kappa}_j^i>\eta_{dual}^i,\,j=0,1,\ldots,N-1\}.
\end{align*}
It follows that sensitivities with larger CMoN values always get updated first. As a consequence, there are finite number of combinations (actually $N+1$) of possible updated sensitivities, resulting in a finite number of possible $P^i$ matrices. Therefore, $\rho^i, \alpha^i, \beta^i$ which are defined in \eqref{rho}, \eqref{alpha} and \eqref{beta}, are functions of $P^i$ and have at most $N+1$ possible values. Hence, the range of $\mathcal{U}_1,\mathcal{U}_2$ are finite. 

In addition, for some $k\neq j$, $\forall (\eta_{pri}^i,\eta_{dual}^i)$ satisfying
\begin{align}\label{eta interval}
\begin{split}
    &\kappa_k^i<\eta_{pri}^i< \kappa_j^i,\\
    &\tilde{\kappa}_k^i<\eta_{dual}^i< \tilde{\kappa}_j^i,
\end{split}
\end{align}
there exist $(\epsilon_{pri},\epsilon_{dual})\neq 0\in\mathcal{R}$ such that 
\begin{align*}
    &\kappa_k^i<\eta_{pri}^i+\epsilon_{pri}< \kappa_j^i,\\
    &\tilde{\kappa}_k^i<\eta_{dual}^i+\epsilon_{dual}< \tilde{\kappa}_j^i.
\end{align*}
Hence, $P^i$ remains constant under the perturbation of $(\epsilon_{pri},\epsilon_{dual})$. The values of $\mathcal{U}_1, \mathcal{U}_2$ are constant for any $(\eta_{pri}^i,\eta_{dual}^i)$ satisfying \eqref{eta interval}. Discontinuity exists at $(\eta_{pri}^i,\eta_{dual}^i)=(\kappa_j^i,\tilde{\kappa}_j^i)$, when the matrix $P^i$ has different number of nonzero blocks resulting in different values of $\mathcal{U}_1, \mathcal{U}_2$. 
\end{IEEEproof}

\bibliographystyle{IEEEtran}
\bibliography{IEEEabrv,journal_paper}

\begin{thebibliography}{10}
\providecommand{\url}[1]{#1}
\csname url@samestyle\endcsname
\providecommand{\newblock}{\relax}
\providecommand{\bibinfo}[2]{#2}
\providecommand{\BIBentrySTDinterwordspacing}{\spaceskip=0pt\relax}
\providecommand{\BIBentryALTinterwordstretchfactor}{4}
\providecommand{\BIBentryALTinterwordspacing}{\spaceskip=\fontdimen2\font plus
\BIBentryALTinterwordstretchfactor\fontdimen3\font minus
  \fontdimen4\font\relax}
\providecommand{\BIBforeignlanguage}[2]{{%
\expandafter\ifx\csname l@#1\endcsname\relax
\typeout{** WARNING: IEEEtran.bst: No hyphenation pattern has been}%
\typeout{** loaded for the language `#1'. Using the pattern for}%
\typeout{** the default language instead.}%
\else
\language=\csname l@#1\endcsname
\fi
#2}}
\providecommand{\BIBdecl}{\relax}
\BIBdecl

\bibitem{bock1984multiple}
H.~G. Bock and K.-J. Plitt, ``A multiple shooting algorithm for direct solution
  of optimal control problems,'' in \emph{Proceedings of the 9th IFAC World
  Congress Budapest, Pergamon, Oxford}, 1984.

\bibitem{biegler2010nonlinear}
L.~T. Biegler, \emph{Nonlinear programming: concepts, algorithms, and
  applications to chemical processes}.\hskip 1em plus 0.5em minus 0.4em\relax
  SIAM, 2010, vol.~10.

\bibitem{powell1978fast}
M.~J. Powell, ``A fast algorithm for nonlinearly constrained optimization
  calculations,'' in \emph{Numerical analysis}.\hskip 1em plus 0.5em minus
  0.4em\relax Springer, 1978, pp. 144--157.

\bibitem{diehl2001real}
M.~Diehl, ``Real-time optimization for large scale nonlinear processes,'' Ph.D.
  dissertation, Heidelberg University, 2001.

\bibitem{zavala2009advanced}
V.~M. Zavala and L.~T. Biegler, ``The advanced-step nmpc controller:
  Optimality, stability and robustness,'' \emph{Automatica}, vol.~45, no.~1,
  pp. 86--93, 2009.

\bibitem{graichen2012real}
K.~Graichen and B.~K{\"a}pernick, \emph{A real-time gradient method for
  nonlinear model predictive control}.\hskip 1em plus 0.5em minus 0.4em\relax
  INTECH Open Access Publisher, 2012.

\bibitem{leineweber1999efficient}
D.~Leineweber, ``Efficient reduced sqp methods for the optimization of chemical
  processes described by large sparse dae models,'' Ph.D. dissertation,
  University of Heidelberg, 1999.

\bibitem{nocedal2006numerical}
J.~Nocedal and S.~Wright, \emph{Numerical optimization}.\hskip 1em plus 0.5em
  minus 0.4em\relax Springer Science \& Business Media, 2006.

\bibitem{martins2003complex}
J.~R. Martins, P.~Sturdza, and J.~J. Alonso, ``The complex-step derivative
  approximation,'' \emph{ACM Transactions on Mathematical Software (TOMS)},
  vol.~29, no.~3, pp. 245--262, 2003.

\bibitem{rall1981automatic}
L.~B. Rall, \emph{Automatic differentiation: Techniques and
  applications}.\hskip 1em plus 0.5em minus 0.4em\relax Springer, 1981.

\bibitem{kuhl2007muscod}
P.~K{\"u}hl, J.~Ferreau, J.~Albersmeyer, C.~Kirches, L.~Wirsching, S.~Sager,
  A.~Potschka, G.~Schulz, M.~Diehl, D.~B. Leineweber \emph{et~al.}, ``Muscod-ii
  users manual,'' \emph{University of Heidelberg}, 2007.

\bibitem{houska2011auto}
B.~Houska, H.~J. Ferreau, and M.~Diehl, ``An auto-generated real-time iteration
  algorithm for nonlinear mpc in the microsecond range,'' \emph{Automatica},
  vol.~47, no.~10, pp. 2279--2285, 2011.

\bibitem{diehl2002real}
M.~Diehl, H.~G. Bock, J.~P. Schl{\"o}der, R.~Findeisen, Z.~Nagy, and
  F.~Allg{\"o}wer, ``Real-time optimization and nonlinear model predictive
  control of processes governed by differential-algebraic equations,''
  \emph{Journal of Process Control}, vol.~12, no.~4, pp. 577--585, 2002.

\bibitem{gros2016linear}
S.~Gros, M.~Zanon, R.~Quirynen, A.~Bemporad, and M.~Diehl, ``From linear to
  nonlinear mpc: bridging the gap via the real-time iteration,''
  \emph{International Journal of Control}, pp. 1--19, 2016.

\bibitem{bock2007constrained}
H.~G. Bock, M.~Diehl, E.~Kostina, and J.~P. Schlo'der, ``Constrained optimal
  feedback control of systems governed by large differential algebraic,''
  \emph{Real-Time PDE-constrained optimization}, vol.~3, p.~1, 2007.

\bibitem{wirsching2006fast}
L.~Wirsching, H.~G. Bock, and M.~Diehl, ``Fast nmpc of a chain of masses
  connected by springs,'' in \emph{Computer Aided Control System Design, 2006
  IEEE International Conference on Control Applications, 2006 IEEE
  International Symposium on Intelligent Control, 2006 IEEE}.\hskip 1em plus
  0.5em minus 0.4em\relax IEEE, 2006, pp. 591--596.

\bibitem{wirsching2008adjoint}
L.~Wirsching, J.~Albersmeyer, P.~K{\"u}hl, M.~Diehl, and H.~Bock, ``An
  adjoint-based numerical method for fast nonlinear model predictive control,''
  in \emph{Proceedings of the 17th IFAC World Congress, Seoul, Korea},
  vol.~17.\hskip 1em plus 0.5em minus 0.4em\relax Citeseer, 2008, pp.
  1934--1939.

\bibitem{kirches2012efficient}
C.~Kirches, L.~Wirsching, H.~Bock, and J.~Schl{\"o}der, ``Efficient direct
  multiple shooting for nonlinear model predictive control on long horizons,''
  \emph{Journal of Process Control}, vol.~22, no.~3, pp. 540--550, 2012.

\bibitem{zanelli2016efficient}
A.~Zanelli, R.~Quirynen, and M.~Diehl, ``An efficient inexact nmpc scheme with
  stability and feasibility guarantees,'' \emph{IFAC-PapersOnLine}, vol.~49,
  no.~18, pp. 53--58, 2016.

\bibitem{chen2017fast}
Y.~Chen, D.~Cuccato, M.~Bruschetta, and A.~Beghi, ``A fast nonlinear model
  predictive control strategy for real-time motion control of mechanical
  systems,'' in \emph{Advanced Intelligent Mechatronics (AIM), 2017 IEEE
  International Conference on}.\hskip 1em plus 0.5em minus 0.4em\relax IEEE,
  2017, pp. 1780--1785.

\bibitem{chen2017inexact}
------, ``An inexact sensitivity updating scheme for fast nonlinear model
  predictive control based on a curvature-like measure of nonlinearity,'' in
  \emph{Decision and Control (CDC), 2017 IEEE 56th Annual Conference on}.\hskip
  1em plus 0.5em minus 0.4em\relax IEEE, 2017, pp. 4382--4387.

\bibitem{van2017towards}
N.~van Duijkeren, G.~Pipeleers, J.~Swevers, and M.~Diehl, ``Towards dynamic
  optimization with partially updated sensitivities,''
  \emph{IFAC-PapersOnLine}, vol.~50, no.~1, pp. 8680--8685, 2017.

\bibitem{albersmeyer2009fast}
J.~Albersmeyer, D.~Beigel, C.~Kirches, L.~Wirsching, H.~G. Bock, and J.~P.
  Schl{\"o}der, ``Fast nonlinear model predictive control with an application
  in automotive engineering,'' in \emph{Nonlinear Model Predictive
  Control}.\hskip 1em plus 0.5em minus 0.4em\relax Springer, 2009, pp.
  471--480.

\bibitem{lindscheid2016parallelization}
C.~Lindscheid, D.~Ha{\ss}kerl, A.~Meyer, A.~Potschka, H.~Bock, and S.~Engell,
  ``Parallelization of modes of the multi-level iteration scheme for nonlinear
  model-predictive control of an industrial process,'' in \emph{Control
  Applications (CCA), 2016 IEEE Conference on}.\hskip 1em plus 0.5em minus
  0.4em\relax IEEE, 2016, pp. 1506--1512.

\bibitem{frasch2012mixed}
J.~V. Frasch, L.~Wirsching, S.~Sager, and H.~G. Bock, ``Mixed---level iteration
  schemes for nonlinear model predictive control,'' \emph{IFAC Proceedings
  Volumes}, vol.~45, no.~17, pp. 138--144, 2012.

\bibitem{schweickhardt2004quantitative}
T.~Schweickhardt and F.~Allgower, ``Quantitative nonlinearity assessment-an
  introduction to nonlinearity measures,'' \emph{Computer Aided Chemical
  Engineering}, vol.~17, pp. 76--95, 2004.

\bibitem{galan2003gap}
O.~Gal{\'a}n, J.~A. Romagnoli, A.~Palazoglu, and Y.~Arkun, ``Gap metric concept
  and implications for multilinear model-based controller design,''
  \emph{Industrial \& engineering chemistry research}, vol.~42, no.~10, pp.
  2189--2197, 2003.

\bibitem{bates1980relative}
D.~M. Bates and D.~G. Watts, ``Relative curvature measures of nonlinearity,''
  \emph{Journal of the Royal Statistical Society. Series B (Methodological)},
  pp. 1--25, 1980.

\bibitem{guay1996measurement}
M.~Guay, ``Measurement of nonlinearity in chemical process control,'' Ph.D.
  dissertation, Queen’s University, 1996.

\bibitem{niu2008curvature}
R.~Niu, P.~K. Varshney, M.~Alford, A.~Bubalo, E.~Jones, and M.~Scalzo,
  ``Curvature nonlinearity measure and filter divergence detector for nonlinear
  tracking problems,'' in \emph{Information Fusion, 2008 11th International
  Conference on}.\hskip 1em plus 0.5em minus 0.4em\relax IEEE, 2008, pp. 1--8.

\bibitem{mallick2005differential}
M.~Mallick and B.~F. La~Scala, ``Differential geometry measures of nonlinearity
  for ground moving target indicator (gmti) filtering,'' in \emph{Information
  Fusion, 2005 8th International Conference on}, vol.~1.\hskip 1em plus 0.5em
  minus 0.4em\relax IEEE, 2005, pp. 219--226.

\bibitem{guay1997measurement}
M.~Guay, P.~McLellan, and D.~Bacon, ``Measurement of dynamic process
  nonlinearity,'' \emph{IFAC Proceedings Volumes}, vol.~30, no.~9, pp.
  589--594, 1997.

\bibitem{li2012measure}
X.~R. Li, ``Measure of nonlinearity for stochastic systems,'' in
  \emph{Information Fusion (FUSION), 2012 15th International Conference
  on}.\hskip 1em plus 0.5em minus 0.4em\relax IEEE, 2012, pp. 1073--1080.

\bibitem{diehl2010adjoint}
M.~Diehl, A.~Walther, H.~G. Bock, and E.~Kostina, ``An adjoint-based sqp
  algorithm with quasi-newton jacobian updates for inequality constrained
  optimization,'' \emph{Optimization Methods \& Software}, vol.~25, no.~4, pp.
  531--552, 2010.

\bibitem{daniel1973stability}
J.~W. Daniel, ``Stability of the solution of definite quadratic programs,''
  \emph{Mathematical Programming}, vol.~5, no.~1, pp. 41--53, 1973.

\bibitem{fiacco1983introduction}
A.~V. Fiacco, \emph{Introduction to sensitivity and stability analysis in
  nonlinear programming.}\hskip 1em plus 0.5em minus 0.4em\relax Academic
  press, 1983.

\bibitem{zavala2010real}
V.~M. Zavala and M.~Anitescu, ``Real-time nonlinear optimization as a
  generalized equation,'' \emph{SIAM Journal on Control and Optimization},
  vol.~48, no.~8, pp. 5444--5467, 2010.

\bibitem{dontchev2013euler}
A.~L. Dontchev, M.~Krastanov, R.~T. Rockafellar, and V.~M. Veliov, ``An
  euler--newton continuation method for tracking solution trajectories of
  parametric variational inequalities,'' \emph{SIAM Journal on Control and
  Optimization}, vol.~51, no.~3, pp. 1823--1840, 2013.

\bibitem{Andersson2013b}
J.~Andersson, ``{A} {G}eneral-{P}urpose {S}oftware {F}ramework for {D}ynamic
  {O}ptimization,'' {P}h{D} thesis, Arenberg Doctoral School, KU Leuven,
  Department of Electrical Engineering (ESAT/SCD) and Optimization in
  Engineering Center, Kasteelpark Arenberg 10, 3001-Heverlee, Belgium, October
  2013.

\bibitem{hpipm}
``Hpipm,'' \url{https://github.com/giaf/hpipm}.

\bibitem{quirynen2015autogenerating}
R.~Quirynen, M.~Vukov, M.~Zanon, and M.~Diehl, ``Autogenerating microsecond
  solvers for nonlinear mpc: A tutorial using acado integrators,''
  \emph{Optimal Control Applications and Methods}, vol.~36, no.~5, pp.
  685--704, 2015.

\bibitem{o2013splitting}
B.~O'Donoghue, G.~Stathopoulos, and S.~Boyd, ``A splitting method for optimal
  control,'' \emph{IEEE Transactions on Control Systems Technology}, vol.~21,
  no.~6, pp. 2432--2442, 2013.

\bibitem{enns2010s}
R.~H. Enns, \emph{It's a nonlinear world}.\hskip 1em plus 0.5em minus
  0.4em\relax Springer Science \& Business Media, 2010.

\bibitem{griewank2002constrained}
A.~Griewank and A.~Walther, ``On constrained optimization by adjoint based
  quasi-newton methods,'' \emph{Optimization Methods and Software}, vol.~17,
  no.~5, pp. 869--889, 2002.

\bibitem{kouzoupis2015block}
D.~Kouzoupis, R.~Quirynen, J.~Frasch, and M.~Diehl, ``Block condensing for fast
  nonlinear mpc with the dual newton strategy,'' \emph{IFAC-PapersOnLine},
  vol.~48, no.~23, pp. 26--31, 2015.

\end{thebibliography}

\end{document}